\documentclass[times,twocolumn,final]{elsarticle}
\usepackage{enumitem}

\usepackage[ruled,linesnumbered]{algorithm2e}
\usepackage{amsmath}
\usepackage{amssymb} % 为了使用 \checkmark
\usepackage{multirow}
\usepackage{multirow}
\usepackage{medima}
\usepackage{framed,multirow}
\usepackage{amssymb}
\usepackage{latexsym}
\usepackage{graphicx}

\usepackage{url}
\usepackage{xcolor}
\usepackage{booktabs} % 用于更美观的表格线
\usepackage{color}
\usepackage{hyperref}

\definecolor{newcolor}{rgb}{.8,.349,.1}

\journal{Medical Image Analysis}

\begin{document}

\verso{Given-name Surname \textit{et~al.}}

\begin{frontmatter}

\title{CAD-Unet: A Capsule Network-Enhanced Unet Architecture for Accurate Segmentation of COVID-19 Lung Infections from CT Images }%

% \title{CAD-Unet: A Capsule Network-Enhanced Unet Architecture for Accurate Segmentation of COVID-19 Lung Infections from CT Images \tnoteref{tnote1}}%
% \tnotetext[tnote1]{This is an example for title footnote coding.}

\author[1]{Yijie Dang}
% \ead{18209519021@163.com }

\author[1,2]{Weijun Ma \corref{cor1}}
\cortext[cor1]{Corresponding author.}
\ead{weijunma_2008@sina.com }
% \fntext[fn1]{This is author footnote for second author.}
\author[3]{Xiaohu Luo}
%% Third author's email
% \ead{weijunma_2008@sina.com (Weijun Ma)}
%\author[2]{Given-name4 \snm{Surname4}}

\author[4]{Huaizhu Wang}
% \ead{wanghz78@qq.com }

\address[1]{School of Information Engineering, Ningxia University, Yinchuan, 750021, Ningxia, China}
\address[2]{Ningxia Key Laboratory of Artificial Intelligence and Information Security for Channeling Computing Resources from the East to the West, Ningxia University, Yinchuan, 750021, Ningxia, China}
\address[3]{School of Mathematics and Computer Science, Ningxia Normal University, Guyuan, 756000, China}
\address[4]{School of Advanced Interdisciplinary Studies, Ningxia University, Zhongwei, 755000,
China}

\received{1 May 2013}
\finalform{10 May 2013}
\accepted{13 May 2013}
\availableonline{15 May 2013}
\communicated{S. Sarkar}

\begin{abstract}
Since the outbreak of the COVID-19 pandemic in 2019, medical imaging has emerged as a primary modality for diagnosing COVID-19 pneumonia. In clinical settings, the segmentation of lung infections from computed tomography images enables rapid and accurate quantification and diagnosis of COVID-19. Segmentation of COVID-19 infections in the lungs poses a formidable challenge, primarily due to the indistinct boundaries and limited contrast presented by ground glass opacity manifestations. Moreover, the confounding similarity among infiltrates, lung tissues, and lung walls further complicates this segmentation task. To address these challenges, this paper introduces a novel deep network architecture, called CAD-Unet, for segmenting COVID-19 lung infections.
\noindent In this architecture, capsule networks are incorporated into the existing Unet framework. Capsule networks represent a novel type of network architecture that differs from traditional convolutional neural networks. They utilize vectors for information transfer among capsules, facilitating the extraction of intricate lesion spatial information. Additionally, we design a capsule encoder path and establish a coupling path between the unet encoder and the capsule encoder. This design maximizes the complementary advantages of both network structures while achieving efficient information fusion.
\noindent Finally, extensive experiments are conducted on four publicly available datasets, encompassing binary segmentation tasks and multi-class segmentation tasks. The experimental results demonstrate the superior segmentation performance of the proposed model. The code has been released at: \url{https://github.com/AmanoTooko-jie/CAD-Unet}.
\end{abstract}


\end{frontmatter}

%\linenumbers

\section{Introduction}
Since the outbreak of the novel coronavirus disease (COVID-19) in 2019, healthcare systems worldwide have faced tremendous pressure. In such a critical situation, timely and accurate diagnosis is crucial for effective patient management. Currently, reverse transcriptase-polymerase chain reaction (RT-PCR) is considered as the gold standard of diagnosing COVID-19 for its high specificity \citep{wang2020detection}. While PCR testing serves as the primary clinical diagnostic test, it is not without certain limitations. Firstly, the implementation of PCR testing poses challenges in resource-limited settings or remote areas due to its requirement for specialized laboratory facilities and skilled personnel, which are often scarce in such environments. Secondly, the turnaround time for PCR results can be relatively long, ranging from a few hours to days, which may hinder timely decision-making and patient management. Finally, a limitation is the potential for false negatives. Factors such as improper sample collection, variations in viral load during different stages of infection, or technical errors in the PCR process, which can contribute to false-negative results, leading missed diagnoses and potential virus transmission \citep{ai2020correlation}. To address these challenges, medical imaging techniques have been extensively utilized as a primary or adjunctive tool \citep{vantaggiato2021COVID}.

Among many medical imaging techniques, computed tomography (CT) imaging has emerged as a valuable tool in the detection and assessment of COVID-19 \citep{kucirka2020variation}, offering detailed insights into the extent and severity of lung involvement. CT scans of COVID-19 patients commonly exhibit distinctive radiological patterns, such as ground-glass opacities (GGOs), consolidations (CONs), and crazy-paving patterns, as well as various other features. GGO, characterized by an increased density in lung regions with a hazy shadow reminiscent of glass transparency, presents as high-opacity areas with blurred margins on CT scans.  CON, characterized by increased density due to inflammatory exudates or hemorrhage filling alveoli, appears as denser, solid white areas, with potential for clear or blurred edges. Crazy-paving patterns, displaying thickened alveolar walls and increased interstitial texture due to edema and inflammation, exhibit an intricate, brick-like interlocking pattern on CT images. This variability poses difficulties for radiologists in accurately segmenting and distinguishing infected areas from normal lung tissue. As a result, the pursuit of automated segmentation techniques for COVID-19 lung CT images has garnered significant attention in recent years. Automated segmentation of lung regions from CT scans plays a crucial role in quantifying the extent of lung involvement, assisting in disease diagnosis, monitoring disease progression, and evaluating treatment response. It enables the extraction of quantitative features, such as lesion volume and density, which can aid in patient risk stratification and treatment planning \citep{lei2020ct,ng2020imaging,pan2020time}.

Over the years, various computerized methods have been proposed for lung segmentation in CT images, ranging from traditional image processing algorithms to advanced machine learning techniques. Deep learning, particularly convolutional neural networks (CNNs), has emerged as a dominant approach for medical image segmentation, demonstrating promising results in various applications.

However, the complex and variable nature of COVID-19 lesions within the lungs presents significant challenges for automatic segmentation using deep learning techniques.  COVID-19, caused by the SARS-CoV-2 virus, manifests in the lungs through diverse and highly variable pathological findings.  Firstly, the size of COVID-19-associated lesions can vary considerably, with some patients exhibiting small, focal areas of infection, while others display widespread, diffuse involvement of multiple lung lobes.  This variability necessitates deep learning models to possess a high degree of flexibility and adaptability to accurately segment these patterns.  Secondly, the shapes and patterns of the lesions also exhibit significant diversity. GGOs typically manifest as round or oval-shaped opacities, whereas CONs may appear as irregular or patchy areas. In some cases, lesions may present with a mixed appearance, combining GGOs with areas of consolidation or fibrosis.  This heterogeneity in lesion morphology poses considerable challenges for deep learning models to consistently and accurately segment the infected regions.  Finally, the location of lesions within the lungs can vary significantly, with some patients having lesions primarily in peripheral regions and others in more central locations.  This spatial variability requires deep learning models to have a comprehensive understanding of lung anatomy and the potential locations of COVID-19 lesions.

In order to tackle the aforementioned challenges, we introduce an innovative deep network architecture called CAD-Unet designed specifically for CT slices. In this architecture, we couple capsule networks into the encoder stage of the Unet architecture to enhance the segmentation performance of the proposed network. Capsule networks can better preserve the relative spatial positional information between features, thereby enhancing their recognition capabilities for the varying sizes, complex shapes, and differentiated lesion locations exhibited by COVID-19 pulmonary lesions. This capability aligns with the fundamental principle of capsule networks, where information is stored and transmitted at the neuron level as vectors rather than scalars \citep{lalonde2018capsules}. These vectors encapsulate details about the spatial attitude of objects, including their locations, scale locations, and orientations between different parts \citep{tran20223dconvcaps}. Furthermore, the implementation of a dynamic routing algorithm in capsule networks facilitates the transfer of information between capsule layers, allowing the model to effectively capture the positional relationships between objects. We conduct comprehensive comparative experiments using four publicly accessible datasets, illustrating the superior performance of the proposed architectural framework in the context of pulmonary infection segmentation.

In summary, the main contributions of this paper are given as follows:
\begin{itemize}
    \item We introduce a dedicated network architecture designed to address the segmentation task of COVID-19 pulmonary CT lesions, with a particular focus on handling complex lesion shapes that are difficult to identify. This architecture integrates both capsule networks and U-Net, enhancing the model's capability to capture and encode critical information in lung CT images.
    \item We innovatively design the integration of capsule networks pathway and U-Net encoder pathway in parallel, coupled through a pathway coupling mechanism. This design not only maximizes the complementary strengths of the two network structures but also allows for efficient information fusion. The model learns and expresses features more comprehensively, excelling particularly in scenarios involving complex lesion shapes and ambiguous boundaries.
    \item We conduct extensive experiments on four publicly available datasets, including both binary segmentation tasks and multi-class segmentation tasks, and compare the proposed architecture against three baseline architectures (Unet \citep{unet}, UNet++ \citep{zhou2018unet++}, Att-Unet \citep{oktay2018attention}) and four state-of-the-art COVID-19 segmentation architectures (binary segmentation: InfNet \citep{fan2020inf}, SCOAT-Net \citep{zhao2021scoat}, nCoVSegNet \citep{liu2021COVID}, PDEAtt-Unet \citep{bougourzi2023pdatt} and D-TrAttUnet \citep{bougourzi2023d}, multi-class segmentation: CopleNet \citep{wang2020noise}, AnamNet \citep{paluru2021anam}, SCOAT-Net \citep{zhao2021scoat}, and D-TrAttUnet). In addition, we also integrate our architecture into nnUNet \citep{Isensee2021} and compare it with the native nnUNet architecture. The experimental results demonstrate that the proposed architecture exhibits strong segmentation performance in both binary and multi-class segmentation, outperforming the strong baselines and state-of-the-art medical image segmentation methods.
\end{itemize}

\section{Related works}
The COVID-19 pandemic has sparked extensive research in the field of medical imaging, particularly in the domain of lung CT lesion segmentation. There are various methods for lung CT lesion segmentation. In this section, we will only discuss segmentation architectures based on Unet and the utilization of capsule networks for object segmentation.
\subsection{Segmentation architecture based on Unet}
The segmentation architecture based on Unet \citep{unet} has found extensive application in the field of medical imaging, particularly for lung CT lesion segmentation tasks. Unet constitutes a classical CNNs architecture characterized by the concurrent presence of an encoder and a decoder. The encoder gradually extracts both low-level and high-level features from the image, while the decoder maps these features back to the image space, yielding intricate segmentation outcomes.

In the context of the COVID-19 pandemic, numerous researchers have adopted the Unet architecture for segmenting lesion regions in lung CT images, aiming to provide precise support for clinical diagnosis and disease monitoring. For instance, \cite{COVID-19unet++} initially applied Unet++ \citep{zhou2018unet++} to COVID-19 lung CT lesion segmentation. Unet++, an extension of the Unet framework, employs a series of nested and densely connected skip pathways between the encoder and decoder subnetworks, further enhancing the semantic relationships between them and achieving improved performance in lung CT lesion segmentation tasks. \cite{zhao2021scoat} reengineered the connection structure of Unet++ to propose SCOAT-Net, introducing a biologically motivated attention learning mechanism. This approach introduces specially designed spatial-wise and channel-wise attention modules, which collaborate to enhance the network's attention learning and consequently improve the segmentation accuracy. Building upon Unet, \cite{bougourzi2023pdatt} developed PDAtt-Unet, a pyramid dual-decoder Att-Unet \citep{attunet} architecture that utilizes pyramid structures and attention gates to maintain global spatial awareness across all encoding layers. During the decoding phase, PDAtt-Unet incorporates two separate decoders employing attention gates to simultaneously segment infections and the lung region.
\subsection{Capsule networks for object segmentation}
Capsule networks, introduced by \cite{capsulenet}, have attracted considerable attention in the domain of object segmentation, owing to their distinctive architecture and capacity to mitigate specific constraints associated with conventional CNNs. Unlike CNNs, which depend on pooling and hierarchical feature extraction, capsule networks endeavor to characterize the hierarchical interdependencies among features by encapsulating them within capsules.

The fundamental innovation of capsule networks lies in their ability to preserve spatial hierarchies and capture viewpoint information of objects. Traditional CNNs often struggle with handling variations in object poses and tend to rely heavily on the spatial pooling of features. Conversely, capsules within a capsule network encode multiple attributes of an object, such as pose, deformation, and texture, while also considering the relationships between these attributes. This makes capsule networks inherently suited for object segmentation tasks, where understanding spatial hierarchies and viewpoints is critical.

However, the original architecture of capsule networks and the dynamic routing algorithm incur significant computational expenses in terms of memory and runtime \citep{lalonde2018capsules}. To address this concern, \cite{lalonde2018capsules} extended the concept of convolutional capsules and redefined the dynamic routing algorithm in two crucial ways. Firstly, children capsules are selectively routed to parents capsules within a localized spatial kernel. Secondly, transformation matrices are shared among all members of the grid within a specific capsule type, but not shared across different capsule types. Subsequently, a multitude of researchers have delved deeper into the exploration of capsule network models for the purpose of object segmentation. For instance, \cite{survarachakan2020capsule} developed the Multi-SegCaps model, EM-routing SegCaps model, and U-Net model capable of segmenting an arbitrary number of classes. These models were applied to the segmentation of individual 2D slices, utilizing two neighboring slices on each side (five in total). \cite{tran20223dconvcaps} proposed the 3DConvCaps model, which employs capsule networks for 3D image segmentation. 3DConvCaps constitutes a 3D encoder-decoder network with a convolutional capsule encoder, facilitating the learning of lower-level features using convolutional layers (short-range attention), while simultaneously modeling higher-level features (long-range dependencies) with capsule layers.

In this paper, the integration of the two techniques has been undertaken with the aim of COVID-19 lung CT lesion segmentation. However, distinct from these previous investigations, the CAD-Unet establishes a path coupling between the Unet-encoder and convolutional capsules, thereby fusing the strengths of convolution and capsules to extract target features. Moreover, the incorporation of attention gates and a dual decoders has been implemented. These modules collectively interact to significantly bolster the overall performance of the CAD-Unet architecture.  Subsequently, a comprehensive elucidation of the proposed network architecture is furnished.
%\subsection{The Abstract}
\section{The proposed approach}
Within this section, an extensive elucidation of the CAD-Unet network architecture, fundamental constituents of the network, and the employed loss functions will be provided.
\subsection{Network architecture}
\begin{figure*}[h] % 开始图片
\centering % 图片居中
\includegraphics[scale=0.32]{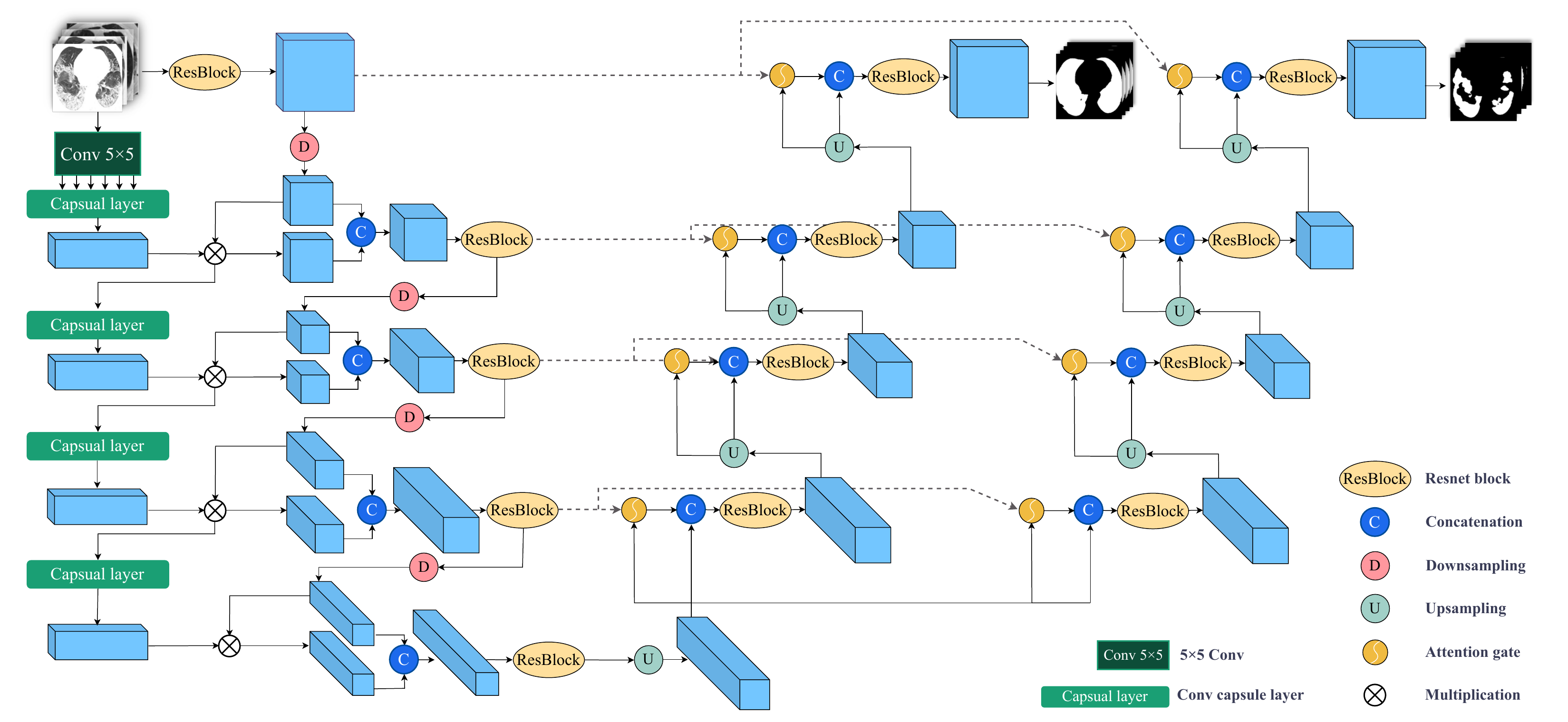} %[设置图像大小]{图片路径(需要在本文件下的路径)}
\caption{Detailed structure of the proposed CAD-Unet architecture.} % 图像描述
\label{fig:1} % 给图片标记
\end{figure*}

The overall architecture of CAD-Unet is illustrated in Fig. \ref{fig:1}. Based on the foundational Unet framework, the capsule encoder path of capsule networks and the encoder path of Unet operate in parallel. Each layer of Unet incorporates ResBlocks to consolidate features. The decoder phase consists of two pathways: one for predicting the infected region, which serves as the final prediction, and the other for predicting the lung region, aiming to focus the network attention on the lung area since the infected region is exclusively present within the lung region. In the following sections, we will provide a detailed description of the core modules and loss functions employed in this network.

% 结束图片

\subsection{Convolutional capsule layer}
Capsule networks consist of multiple capsule layers and innovatively represent the information transfer between capsule layers as vectors rather than scalars. These vectors not only encode specific entity types but also describe how entities are instantiated, including their postures, textures, deformations, and the presence of these features themselves \citep{choi2019attention}. Among them, the norm of the vector indicates the probability of the entity's existence, while the orientation of the vector indicates the configuration of the entity. Information between capsule layers is transmitted through a dynamic routing algorithm, which iteratively updates the weights between capsules, enabling higher-level capsules to effectively route the outputs of lower-level capsules. As a result, more accurate and robust feature representations are obtained. This dynamic routing mechanism is a key characteristics of capsule networks, allowing the network to learn hierarchical representations and spatial relationships of objects. \par
\begin{figure*}[h] % 开始图片
\centering % 图片居中
\includegraphics[scale=0.3]{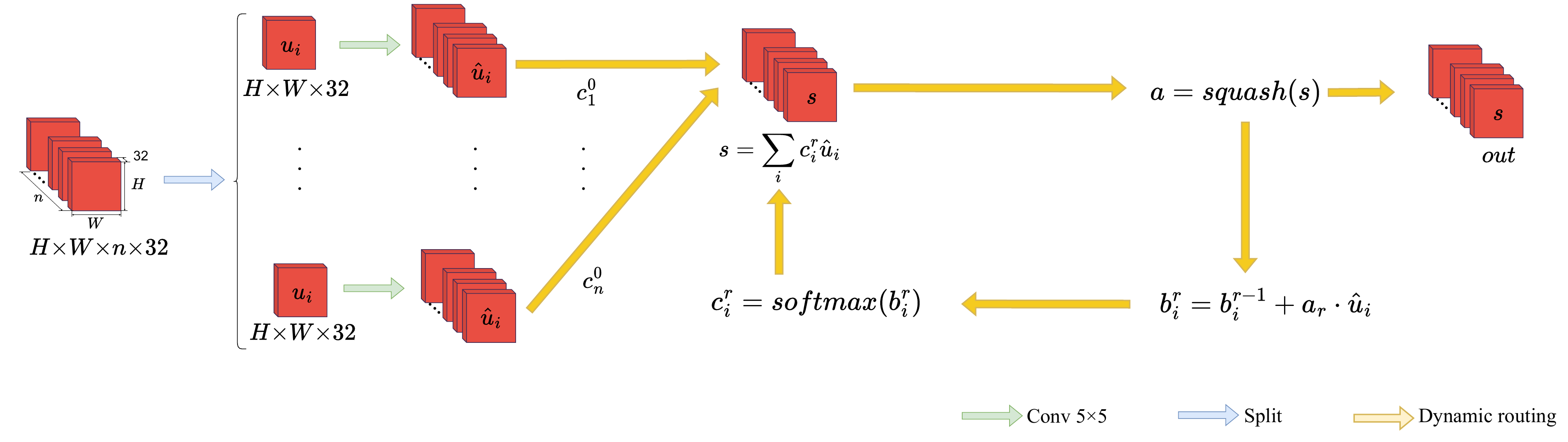} %[设置图像大小]{图片路径(需要在本文件下的路径)}
\caption{Schematic description of convolutional capsule layer} % 图像描述
\label{fig:2} % 给图片标记
\end{figure*}

  Using capsule-based networks for object segmentation poses several issues. The inherent complexity of the original capsule network architecture and dynamic routing algorithm results in significant computational demands, affecting memory usage and runtime performance \citep{lalonde2018capsules}. To address these issues, \cite{lalonde2018capsules} proposed SegCaps, which introduces capsule sharing for each category, effectively reducing the computational overhead. Specifically, as illustrated in Fig.\ref{fig:2}, the input feature map size of the capsule layer is $H\times W\times C\times A$, where $H\times W$ represents the feature map dimensions, $C$ is the number of capsule types, and $A$ is the size of each capsule. Firstly, the capsule $u_i$ for class $i$ is linearly mapped to a higher-level feature $\hat{u}_i$ using the same transformation matrix $W$. Then, all $\hat{u}_i$ are weighted and summed to obtain the input $s$ of the higher-level capsule. Finally, the squash function is applied to $s$ to ensure that the vector direction remains unchanged while compressing the vector length between 0 and 1, representing the probability of entity presence. The dynamic routing process can be represented (\ref{eq:1}) as follows:
\begin{equation}
     \begin{aligned}
     &{\hat{u}}_{i}={W}_{i}{u}_{i} , \\
     &{s}=\sum_{i}c_{i}{\hat{u}}_{i}  ,\\
     &{v}={squash}({s})  ,\\
     &{squash}({s})={\frac{||{s}||^{2}}{1+||{s}||^{2}}}{\frac{{s}}{||{s}||}},
     \label{eq:1}
     \end{aligned}
     \end{equation}
where $i$ is the capsule category, $u$ represents the input to the higher-level capsule, $W$ is the transformation matrix, $\hat{u}_i$ corresponds to the input of the higher-level capsule, $c_i$ represents the coupling coefficients determined through the iterative dynamic routing process, and $v$ denotes the output of the capsule at this layer. $||\cdot||$ denotes $L_2$ norm.  \par Linear mapping is achieved by the convolutional layer, where $W$ actually represents the parameters of the convolutional kernels.   The convolutional layer employs a kernel size of $5\times5$ with a stride of 2, which enables the completion of one downsampling operation on the original feature map.   \par The coupling coefficients $c_i$ between capsule categories are determined by the dynamic routing algorithm.   This algorithm is an iterative process that updates the values of $c_i$ through multiple rounds of iterations.   The number of iterations $r$ is a hyperparameter and is set to 3 in our architecture.

%%[t]：插入到页面的顶部（top）。
%%[b]：插入到页面的底部（bottom）。
%%[h]：尽量将图片或表格放置在原位置（here）。
%%[p]：将图片或表格放置在单独的一页（page）。
% 结束图片
\subsection{Two parallel encoder pathways}
COVID-19 lung lesions exhibit diverse morphologies in CT images, including infiltrations, nodules, and patchy patterns, among others. Solely utilizing convolutional networks for learning target region features has limitations, while capsule networks utilize vectors to convey information, enabling a better capture of target morphology. Additionally, the dynamic routing algorithm connects different layers of capsules, allowing for a better capture of contextual relationships between objects. This advantage empowers capsule networks to handle issues such as object relationships and hierarchical structures in images. Therefore, we innovatively couple capsule networks with convolution in parallel, leveraging the strengths of capsule networks to compensate for the limitations of convolution. \par
\begin{figure}[h] % 开始图片
\centering % 图片居中
\includegraphics[scale=0.45]{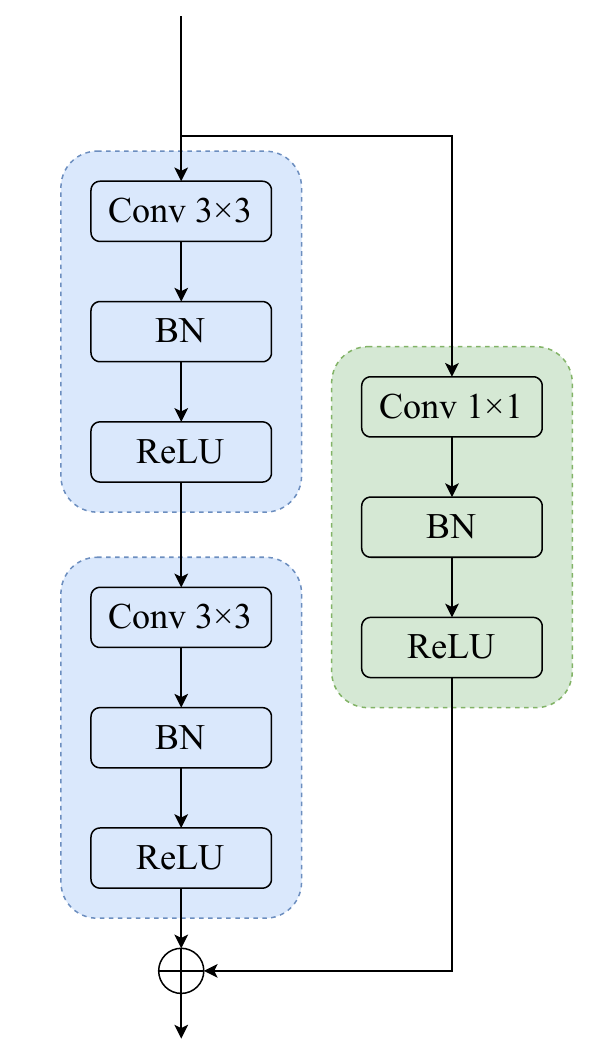} %[设置图像大小]{图片路径(需要在本文件下的路径)}
\caption{The structure of ResBlock} % 图像描述
\label{fig:4} % 给图片标记
\end{figure}

 Specifically, in the encoder stage of the network, we  design two parallel paths: the capsule encoder path and the Unet encoder  path, and the feature information from these two paths interacts with each other. The input image $x\in \mathbb{R}^{H\times W\times C}$ is fed into both paths, where $H$ and $W$ represent the height and width of the input image, respectively, and $C$ denotes the number of channels in the input image. Now, we will provide a detailed explanation of each path.
\begin{enumerate}[label=(\alph*)]
     \item\textbf{Capsules encoder pathway:} In the capsule encoding pathway, we initially generate primary capsules through a convolutional layer with a kernel size of $5\times 5$, resulting in capsules of a single category $C_0 \in \mathbb{R}^{H\times W\times 1\times 16}$.The primary capsule $C_0$ is then propagated to the primary capsule layer, which is a conventional convolutional capsule layer. It undergoes one routing iteration and returns 16-dimensional capsules for two categories, denoted as $C_1 \in \mathbb{R}^{\frac{H}{2}\times \frac{W}{2}\times 2\times 16}$. Following this, at each stage, the convolutional capsule layer is coupled with the results of different levels of the Unet pathway, gradually extracting higher-level semantic information, allowing the model to effectively comprehend the complex structure of input data. Specifically, $C_1$ is coupled with the first layer of the Unet pathway, the coupled result is then forwarded to the secondary capsule layer, comprising three routing iterations. The output consists of 32-dimensional capsules for two categories, denoted as $C_2 \in \mathbb{R}^{\frac{H}{4}\times \frac{W}{4}\times 2\times 32}$, with capsule dimensions expanding similarly through convolutional capsule layer operations. Following this, $C_2$ is coupled with the results of the Unet pathway and transmitted to the tertiary capsule layer, returning 32-dimensional capsules for four categories, represented as $C_3 \in \mathbb{R}^{\frac{H}{8}\times \frac{W}{8}\times 4\times 32}$. Finally, $C_3$ is coupled with the Unet pathway results and transmitted to the quaternary capsule layer, yielding 64-dimensional capsules for four categories, denoted as $C_4 \in \mathbb{R}^{\frac{H}{14}\times \frac{W}{14}\times 4\times 64}$. It is evident that the size of the feature map is halved, while the number of channels is doubled after the convolutional capsule layer. The convolutional operations within the capsule layer regulate the size of the output capsules, with a stride of 2 employed to achieve twofold downsampling. This strategy reduces the spatial dimensions and incrementally increases the channel numbers in the feature map, thereby enabling the model to learn abstract hierarchical features. This process is formalized in equation (\ref{eq:2}):

     \begin{equation}
     \begin{aligned}
     C_0&=Conv_{5\times5}(x_{in}) , \\
     C_1&=ConvCap(C_0) ,\\
     C_2&=ConvCap(C_1\cdot out_0) ,\\
     C_3&=ConvCap(C_2\cdot out_1) ,\\
     C_4&=ConvCap(C_3\cdot out_2),
     \label{eq:2}
     \end{aligned}
     \end{equation}
     where $Conv_{5\times5}(\cdot)$ is the convolution operation with a $5\times5$ kernel, $ConvCap(\cdot)$ is the convolutional capsule layer, and $x_{in}$ is the initial input to the network.
    \item\textbf{Unet encoder pathway:} In the Unet encoder pathway, we employ a Unet baseline architecture enhanced with ResBlocks \citep{bougourzi2023d}. This architecture comprises multiple layers, each consisting of a ResBlock and a downsampling module. The initial layer (Layer 0) receives the original network input, while subsequent layers receive the output of the previous layer and the pathway coupling results from the capsules pathway. Specifically, we perform point-wise multiplication between the predictions from each layer of the capsules pathway and the feature extraction from each layer of the Unet pathway. This strategic fusion enables the Unet pathway to seamlessly integrate features extracted from the capsules pathway, thereby enriching its contextual understanding. Let $out_{l-1}$ represent the output of the previous layer, and the input to the current layer, denoted as $in_l$, is computed as $in_l=cat(out_{l-1}, out_{l-1} \cdot C_l)$. 
    Here, $in_l$ signifies the input to the $l$-th layer, $out_{l-1}$ is the output of the ($l-1$)-th layer, $C_l$ represents the output of the $l$-th capsule layer, and $Cat(\cdot)$ denotes the concatenation operation along the channel dimension. Following this, this input undergoes a ResBlock to extract features, followed by passage through a max-pooling layer for downsampling, resulting in the output of this layer, as depicted in equation (\ref{eq:3}):
    \begin{equation}
     \begin{aligned}
     {in}_0 &=ResBlock(in),~out_0=MP(in_0) , \\
     in_{1} &=Cat(C_1\cdot out_0,out_0),~out_1=MP(RB(in_1)) ,\\
     in_{2} &=Cat(C_2\cdot out_1,out_1),~out_2=MP(RB(in_2)) ,\\
     in_{3} &=Cat(C_3\cdot out_2,out_2),~out_3=MP(RB(in_3)) ,\\
     {in}_4 &=Cat(C_4\cdot out_3,out_3),~out_4=RB(in_4),
     \label{eq:3}
     \end{aligned}
     \end{equation}
    where $in$ is the initial input to the network, $in_l$ is the input of the $l$-th layer, $out_l$ is the output of the $l$-th layer, $C_l$ is the output of the $l$-th capsule layer, $Cat(\cdot)$ denotes the concatenation operation along the channel dimension, $MP(\cdot)$ represents passing through a max-pooling layer, and $RB(\cdot)$ represents passing through a ResBlock. As shown in Fig.\ref{fig:4}, the ResBlock consists of two $3\times 3$ convolutional blocks, each followed by batch normalization and ReLU activation. Additionally, a residual connection is used to add the input to the output of the two convolutional layers. The residual connection consists of a $1\times 1$ convolutional block, followed by batch normalization and ReLU activation as shown in equation (\ref{eq:4}):
    \begin{equation}
     \begin{aligned}
     RB(x)&=Out_{Conv}(x)+Out_{Res}(x) , \\
     Out_{Conv}(x)&=ConvBlock(ConvBlock(x)) ,\\
     Out_{Res}(x)&=ReLU(BN(Conv_{1\times1}(x))) ,\\
     ConvBlock(x)&=ReLU(BN(Conv_{3\times3}(x))) ,
     \label{eq:4}
     \end{aligned}
     \end{equation}
    where $x$ is the input to the ResBlock, $Conv_{n\times n}(\cdot)$ is the convolution operation with an $n\times n$ kernel, $BN(\cdot)$ is the batch normalization, and $ReLU(\cdot)$ is the ReLU activation function.
\end{enumerate}
% 结束图片
\subsection{Attention gate}
The conventional Unet framework incorporates skip connections that directly link low-level features from the encoder to high-level features in the decoder. However, this strategy introduces redundant and irrelevant information to the high-level features, adversely impacting the precision of target segmentation. Motivated by the Attention U-Net \citep{oktay2018attention}, we integrate attention gates into the proposed architecture to address this limitation. The attention gates selectively emphasize relevant features while suppressing redundant and less informative ones, which greatly improves segmentation accuracy. By adaptively weighting the contributions of encoder and decoder features, the attention mechanism enhances the model's capability to focus on crucial regions, thereby refining the segmentation outcomes.\par
\begin{figure*}[h] % 开始图片
\centering % 图片居中
\includegraphics[scale=0.5]{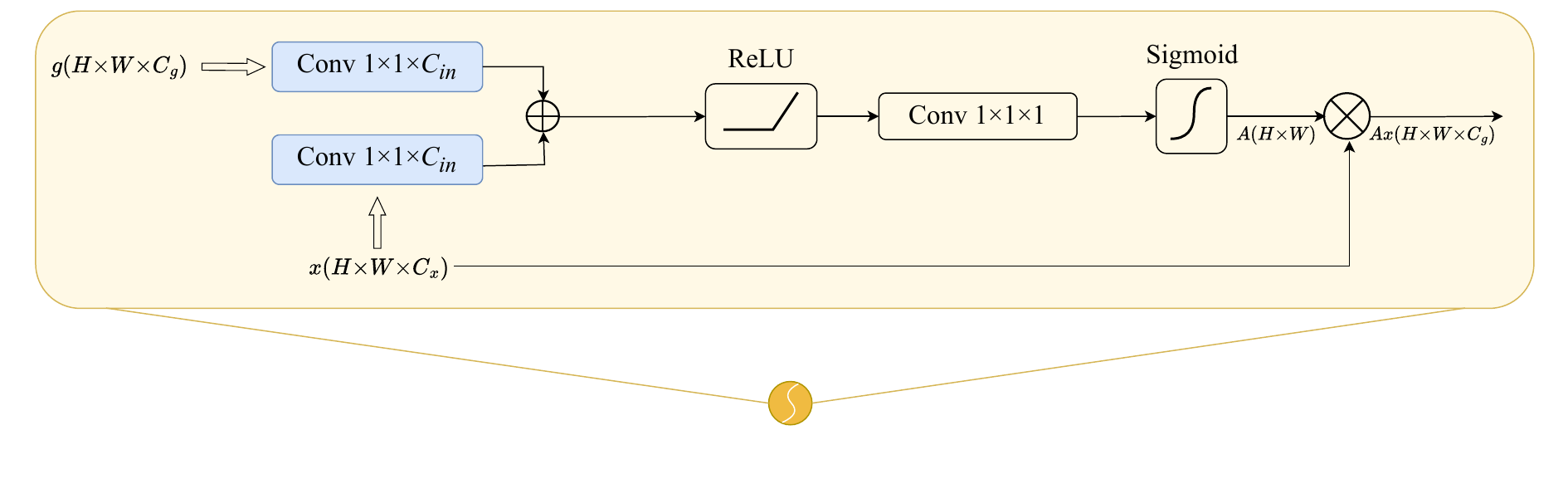} %[设置图像大小]{图片路径(需要在本文件下的路径)}
\caption{Attention gate block, where $g$ is the gating signal and $x$ is the input feature maps. $A$ is the spatial attention obtained, which is applied to all channels of the input feature maps $x$.} % 图像描述
\label{fig:5} % 给图片标记
\end{figure*}

 The structure of the attention gate is elucidated in Fig.\ref{fig:5}. This module processes an input feature map $x\in R^{H\times W\times C_x}$ and a gate signal $g\in R^{H\times W\times C_g}$, where $x$ signifies the output feature map of the encoder $l$-th layer, and $g$ represents the output feature map of the decoder $(l+1)$-th layer. Subsequently, the feature map $x$ and the gate signal $g$ undergo linear transformations through $1\times1$ convolutional layers, followed by batch normalization and element-wise summation. Following this, the ReLU activation function is applied, and another $1\times1$ convolutional layer is utilized to derive spatial attention coefficients, denoted as $Att$ for each pixel. Finally, these obtained attention coefficients are applied to the input feature map $x$, as expressed in equation (\ref{eq:7}):
  \begin{equation}
     \begin{aligned}
     &Att=\psi_{i}(ReLU(BN(W_{x}x_{i})+BN(W_{g}g_{i})))\oplus F , \\
     &out_{AG}=x_{i}\cdot Att ,
     \label{eq:7}
     \end{aligned}
  \end{equation}
where $W_x$ and $W_g$ are $1\times 1$ convolutional operations used for linear transformations, and $\psi_{i}$ is a $1\times 1$ convolutional operation employed to generate spatial attention coefficients. $out_{AG}$ represents the output of the attention gate module.
% 结束图片

\subsection{Dual decoders}
To focus the network on the infected regions, we employ the dual decoders used in DAtt-Unet \citep{bougourzi2023pdatt}. The dual decoders is designed to achieve simultaneous segmentation of both infections regions and lung regions. The objective is to guide the training process to explore the interior of the lung regions and differentiate between infected tissue and non-pulmonary tissue. \par As shown in Fig.\ref{fig:1}, the decoder consists of two pathways: pathway 1 predicts the lung region, and pathway 2 predicts the infection region. The prediction results from pathway 2 serve as the final prediction target, while the existence of pathway 1 aims to guide the model to focus on the lung region, as infection areas only occur within the lungs.

Specifically, except for the last layer, each layer in the decoder receives two inputs: the output from the previous layer of the decoder and the concatenated result of the attention gate output at the same layer. Similar to the encoder, these inputs undergo a ResBlock and an upsampling operation to produce the current layer's output. This output serves two purposes: 1) it is forwarded to the attention gate unit as the gating signal $g$, and 2) it is passed to the next layer of the decoder. The dual decoders process can be represented as equation (\ref{eq:8}):
\begin{equation}
\begin{aligned}
    &d_{i,j} = RB_j(Cat_j(AG_j(US_j(d_{i-1,j}),x_{i,j}),US_j(d_{i-1,j}))), \\
    &\quad \quad \quad \quad \quad \quad i\in \{0,1,2,3,4\},j\in\{1,2\},
    \label{eq:8}
\end{aligned}
\end{equation}
where $i$ represents the layer index, and $j$ represents the decoder pathway number. $d_i$ denotes the output of the $i$-th layer of the decoder, and $x_i$ represents the output of the $i$-th layer of the encoder. The upsampling operation is denoted by US, and AG represents the attention gate unit. It is essential to note that the two pathways of the decoder do not share any modules, including the attention gate unit.
\subsection{Hybrid loss function}
For CT lung lesion segmentation tasks, the complex morphology of the target region edges poses issue for accurate segmentation. Existing works have shown that edge information can provide valuable constraints to guide feature extraction for segmentation \citep{zhao2019egnet,wu2019stacked,zhang2019net}. Therefore, we employ the hybrid loss function proposed in PDEAttUnet \citep{bougourzi2023pdatt} to guide the network training, which consists of three components: 1) infection segmentation loss ($\mathcal{L}_{Inf}$) for the final task prediction, 2) lung segmentation loss ($\mathcal{L}_{Lung}$) to assist in the final prediction of lung regions, and 3) edge loss ($\mathcal{L}_{Edge}$) to enhance the segmentation of target edges. The hybrid loss function is formulated as equation (\ref{eq:10}):
\begin{equation}
\begin{aligned}
    &\mathcal{L}_{All}=\alpha\mathcal{L}_{Inf}+\beta\mathcal{L}_{Lung}+\gamma\mathcal{L}_{Edge}, \\
    &\mathcal{L}_{Inf}=-\sum_{m=1}^{B}\sum_{i=1}^{W\cdot H}G_{R_{i}}\log(p_{i})+(1-G_{R_{i}})\log(1-p_{i}),\\
    &\mathcal{L}_{Lung}=-\sum_{m=1}^{B}\sum_{i=1}^{W\cdot H}G_{L_{i}}\log(s_{i})+(1-G_{L_{i}})\log(1-s_{i}),\\
    &{\cal L}_{Edge}=-\sum_{m=1}^{B}\sum_{i=1}^{W\cdot H}G_{E_{i}}\log(P_{E_{i}})+(1-G_{E_{i}})\log(1-P_{E_i}),
    \label{eq:10}
\end{aligned}
\end{equation}
where $\alpha$, $\beta$, and $\gamma$ are the weights for the infection segmentation loss ($\mathcal{L}_{\text{Inf}}$), lung segmentation loss ($\mathcal{L}_{\text{Lung}}$), and edge loss ($\mathcal{L}_{\text{Edge}}$), respectively. $W$ and $H$ represent the width and height of the predicted mask, and $B$ is the batch size. The ground truth labels for infection and lung of pixel $i$ are denoted as $G_{R_i} \in \{0, 1\}$ and $G_{L_i} \in \{0, 1\}$, respectively. Additionally, $p_i$ and $s_i$ represent the prediction probabilities of infection and lung for pixel $i$, obtained from the decoders used for segmenting infection and lung, respectively. On the other hand, foredge loss function, $G_{E_i} \in \{0, 1\}$ and $P_{E_i}$ are the ground truth label and the predicted probability of edge infection for pixel $i$, respectively. The ground-truth edge pixels are derived by applying morphological gradient to the ground-truth infection regions. Once the ground-truth edge pixels are obtained, the binary cross entropy (BCE) loss function is employed to compute the loss between the ground-truth edge pixels and their corresponding pixels in the mask map. The specific values of the weights $\alpha$, $\beta$, and $\gamma$ will be described in the experimental setup section.
\begin{table*}[h]
\centering
\begin{tabular}{cccc}
\hline
Name     & Dataset                      & CT-Scans & Slices \\ \hline
Dataset 1 & COVID-19 CT segmentation \citep{Radiologists2019}     & 40       & 100    \\
Dataset 2 & Segmentation dataset nr. 2 \citep{Radiologists2019}   & 9        & 829    \\
Dataset 3 & COVID-19-CT-Seg dataset \citep{ma2021toward}      & 20       & 3520   \\
Dataset 4 & CC-CCII segmentation dataset \citep{zhang2020clinically} & 150      & 750    \\ \hline

\end{tabular}
\caption{The used datasets.}
\label{tab:1}
\end{table*}

\begin{table*}[]
\centering
\begin{tabular}{cccccc}
\hline
 Name     &  Total slices &  Non-infected &  Mild &  Intermediate &  Severe \\ \hline
 Dataset 1 & 100          & 1            & 52   & 27           & 20     \\
 Dataset 2 & 829          & 457          & 280  & 48           & 44     \\
 Dataset 3 & 1844         & 0            & 1582 & 193          & 69     \\
 Dataset 4 & 750          & 201          & 462  & 72           & 15     \\ \hline
\end{tabular}
\caption{Distribution of CT slices by infection severity levels across the datasets.}
\label{tab:a}
\end{table*}
\section{Experiments and results}
\subsection{Dataset}
The proposed method is evaluated on four publicly available datasets, namely COVID-19 CT segmentation \citep{Radiologists2019}, segmentation dataset nr.2 \citep{Radiologists2019}, COVID-19-CT-Seg dataset \citep{ma2021toward}, and CC-CCII segmentation dataset  \citep{zhang2020clinically}. Table \ref{tab:1} provides a summary of these datasets.

These datasets will be used to evaluate the segmentation performance of the proposed architecture on binary and multi-class segmentation tasks. In the binary segmentation, the labels are divided into two classes: background and infection region. In the multi-class segmentation, the labels are divided into three classes: background, GGO, and CON.

The COVID-19 CT Segmentation dataset \citep{Radiologists2019} comprises 100 axial CT slices collected from over 40 patients diagnosed with COVID-19. This dataset provides multi-class segmentation labels and is intended for use in multi-class segmentation tasks.

Segmentation dataset nr.2 \citep{Radiologists2019} includes 9 3D CT scans, comprising a total of 829 slices. Of these, 373 slices have been annotated by radiologists with both binary and multi-class segmentation labels for COVID-19 infection. This dataset will be utilized for binary and multi-class segmentation tasks.

The COVID-19-CT-Seg dataset \citep{ma2021toward} includes 20 CT scans from confirmed COVID-19 cases.  It covers a wide range of infection percentages (from 0.01\% to 59\%). A total of 1844 CT slices are annotated with binary infection labels, a meticulous process carried out by a multi-tiered team ranging from junior annotators to senior radiologists. This dataset is intended for binary segmentation tasks.

The CC-CCII segmentation dataset \citep{zhang2020clinically} includes 750 CT slices from 150 COVID-19 patients, with manual annotations for background, lung field, GGO, and CON segmentation. This dataset will be employed for both binary and multi-class segmentation tasks.

Furthermore,to evaluate the quality and potential biases of the datasets used in this study, we provide a detailed breakdown of the distribution of infection severity levels across the CT slices in each dataset. Table \ref{tab:a} summarizes the number of CT slices categorized by different infection severity levels, as defined by \cite{zhang2020clinically}.: mild, defined as less than three ground-glass opacity (GGO) lesions of size less than 3 cm; intermediate, defined as a lesion area exceeding 25\% of the entire lung field; and severe, defined as a lesion area exceeding 50\% of the entire lung field. As observed in the table, non-infected slices in dataset 2 account for more than half of the data (55\%). To mitigate the impact of this bias on model training and validation, we performed random undersampling to reduce their proportion to approximately 30\%. This adjustment helps prevent the model from being skewed towards the majority class (non-infected slices) and improves its ability to learn meaningful features from infected cases. It is important to note that the random undersampling process was carefully implemented to preserve the overall diversity and representativeness of the dataset.

\subsection{Evaluation metrics}
The model is evaluated by  using the following metrics.
\begin{enumerate}

    \item \textbf{F1 score (F1-S):} The F1 score is the harmonic mean of precision and recall, indicating the balance between the true positive rate and the false positive rate,
    \begin{equation}\mathrm{F1score}=100 \cdot{ \frac{2 \cdot TP}{2 \cdot TP+FP+FN}}.\label{eq:11}\end{equation}
    \item \textbf{Intersection over union (IoU):} The IoU, also referred to as the Jaccard index, computes the ratio of the intersection to the union of predicted and ground truth regions,
    \begin{equation}IoU=100\cdot{\frac{TP}{TP+FP+FN}}.\label{eq:12}\end{equation}
    \item \textbf{Recall(Rec):} The Recall, also referred to as sensitivity or the true positive rate, quantitatively measures the proportion of true positive predictions among all actual positive instances,
    \begin{equation}Rec=100\cdot\frac{TP}{TP+FN}.\label{eq:13}\end{equation}
    \item \textbf{Specificity (Spec):} The Specificity measures the proportion of true negative predictions among all actual negative instances,
    \begin{equation}Spec=100\cdot\frac{TN}{FP+TN}.\label{eq:14}\end{equation}
    \item \textbf{Precision (Prec):} The Precision represents the proportion of true positive predictions among all positive predictions made by the model,
    \begin{equation}Prec=100\cdot\frac{TP}{TP+FP},\label{eq:15}\end{equation}
\end{enumerate}
where $TP$, $TN$, $FP$, $FN$ are True Positives, True Negatives, False Positives, False Negatives, respectively.

\subsection{Experimental setup}
We conduct model training using Python 3.7.12 and the PyTorch 1.13.0 framework. The experimental platform consists of an Intel Core i5 13600KF CPU, a single NVIDIA GeForce RTX 3090 Ti 24GB GPU, and the Windows 11 operating system. The libraries used include NumPy 1.21.6, OpenCV 4.7.0, MONAI 1.1.0, among others. The Adam optimizer is adopted for training, with a batch size of 6 and a total of 120 epochs. The initial learning rate is set to 1e-4 and is decayed by a factor of 0.1 after the 50th epoch and again after the 90th epoch. Additionally, we apply data augmentation to the dataset, including random angle rotations ranging from -35 degrees to 35 degrees, as well as random horizontal and vertical flips. In terms of the hybrid loss function, the weights $\alpha$, $\beta$, and $\gamma$ are set to 0.7, 0.3, and 1, respectively. The main experimental settings are summarized in Table \ref{tab:b}.
\begin{table}[]
\begin{tabular}{ll}
\hline
{ Category}             & { Specification}                                                                      \\ \hline
{ CPU}                  & { Intel Core i5 13600KF}                                                              \\
{ GPU}                  & { \begin{tabular}[c]{@{}l@{}}NVIDIA GeForce RTX 3090 Ti\\ 24GB\end{tabular}}          \\
{ Operating System}     & { Windows 11}                                                                         \\
{ RAM}                  & { 32GB DDR4}                                                                          \\
{ Programming Language} & { Python 3.7.12}                                                                      \\
{ Framework}            & { PyTorch 1.13.0}                                                                     \\
{ CUDA Version}         & { 11.7}                                                                               \\
{ cuDNN Version}        & { 8.5.0}                                                                              \\
{ Additional Libraries} & { \begin{tabular}[c]{@{}l@{}}NumPy 1.21.6, OpenCV 4.7.0, \\ MONAI 1.1.0\end{tabular}} \\
{ Epoch}                & { 120}                                                                                \\
{ Batchsize}            & { 6}                                                                                  \\
{ Optimizer}            & { Adam}                                                                               \\
{ Learning rate}        & { 1E-04}                                                                              \\ \hline
\end{tabular}
\caption{Summary of experimental setup.}
\label{tab:b}
\end{table}

\subsection{Experimental results}
In this study, we conduct comprehensive experiments to evaluate the proposed model for both binary and multi-class segmentation tasks of COVID-19 lung CT lesions. All the parameters of the compared methods are set according to the recommendations in the original references or adjusted to achieve the best visual and quantitative results. In addition, nnUNet is widely recognized as a robust baseline for medical image segmentation. To further assess the effectiveness of our approach, we incorporate the proposed architecture into nnUNet, which we term nnCAD-UNet, and compare its performance with the original nnUNet on both segmentation tasks. All experimental results are reported as the mean ± standard deviation over five independent runs. The following sections provide a qualitative analysis of the two tasks.
\subsubsection{Binary segmentation task}
For the binary segmentation task, we compare the proposed model with U-Net, U-Net++, Att-UNet, InfNet, SCOAT-Net, nCoVSegNet, PDEAtt-UNet, and D-TrAttUnet. To evaluate the impact of the hybrid loss on the segmentation performance of different models, we conducted experiments by incorporating the hybrid loss into U-Net, U-Net++, Att-UNet, SCOAT-Net, and nCoVSegNet (note that InfNet, PDEAtt-UNet, and D-TrAttUnet use their own specific loss functions). The experimental results on datasets 2, 3, and 4 are presented in Tables \ref{tab:2}, \ref{tab:3}, and \ref{tab:4}.

On dataset 2, the proposed model outperforms all compared models across multiple metrics, including nnUNet and nnCAD-UNet. Specifically, compared to the best-performing dedicated COVID-19 segmentation model, nCoVSegNet with hybrid loss, the proposed model improves the F1 score, IoU, and Recall by 4.25, 3.65, and 5.9, respectively. When compared to nnUNet, the overall best-performing model among all methods, the proposed model achieves improvements of 1.87, 0.54, and 10.56 in these three metrics. However, integrating the proposed model into nnUNet leads to performance degradation across all three metrics, likely due to the specific infection characteristics in dataset 2 and the relatively large proportion of non-infection CT slices, which may not align well with nnUNet’s preprocessing strategy.

Compared to dataset 2, dataset 3 exhibits overall improved segmentation performance, with F1 scores surpassing 70\% for most models. The general improvement in IoU and Recall, along with lower standard deviations, suggests that the lesions in dataset 3 are more distinct, making segmentation relatively less challenging. On this dataset, the proposed model continues to demonstrate strong segmentation performance. Specifically, compared to the best-performing model among the compared methods, AttUNet, the proposed model achieves improvements of 1.33, 1.81, and 10.12 in F1 score, IoU, and Recall, respectively. Additionally, nnUNet, with its automatically configured architecture, continues to exhibit strong general segmentation capabilities, outperforming all non-nnUNet-based models. However, integrating the proposed model into nnUNet results in improvements of 1.15 and 1.64 in F1 score and IoU, respectively, compared to the original nnUNet, while the Recall metric decreases by 2.51. This indicates that the integration may enhance certain aspects of segmentation accuracy but could compromise the model's ability to fully capture true positive cases.

Among the three datasets, dataset 4 yields the highest segmentation performance across all models. The F1 scores range from 71.61\% (U-Net) to 85.56\% (nnCAD-UNet), with IoU values also being the highest among the datasets.  Specifically, the proposed model achieves the best results, outperforming SCOATNet, the top-performing model among non-nnUNet-based architectures, by 1.15, 1.6, and 9.4 in terms of F1 score, IoU, and Recall respectively.  After integrating this architecture into nnUNet, compared to the original nnUNet, improvements are observed in all three metrics. This indicates that, whether in its original form or when integrated into nnUNet, our proposed Capsule Network-Enhanced Unet Architecture exhibits excellent adaptability and robustness.

In addition, analysis of the experimental results comparing models with and without hybrid loss across all datasets reveals that while the F1 score remains relatively unchanged or experiences a slight decrease, the Recall on most models improves. This aligns with the priorities in medical diagnostics, where minimizing false negatives is of utmost importance. A higher Recall ensures the identification of more true positive cases, thus reducing the risk of missed diagnoses. Although the F1 score does not show a substantial increase, the enhanced Recall improves the model's ability to detect lesions, which is critical for early COVID-19 detection and diagnosis. Therefore, we conclude that the hybrid loss function provides a valuable improvement for the binary segmentation task of COVID-19 lesions.

\begin{table*}[h]
\centering
\begin{tabular}{cccccc}
\hline
Model                                          & F1 Score                                    & IoU                                         & Recall                                      & Spec                                        & Prec                                        \\ \hline
Unet                                           & { 45.23±13.00}          & { 30.15±11.05}          & { 37.88±12.63}          & { 99.39±0.45}           & { 60.47±16.54}          \\
UNet++                                         & { 52.50±9.96}           & { 36.23±9.42}           & { 43.54±11.96}          & { 99.55±0.23}           & { 69.67±9.96}           \\
Att-Unet                                       & { 63.63±8.74}           & { 46.77±6.02}           & { 48.96±7.37}           & { 99.89±0.13}           & { 91.41±7.22}           \\
InfNet                                         & { 47.96±5.30}           & { 31.69±4.40}           & { 42.92±4.78}           & { 99.17±0.38}           & { 55.54±11.26}          \\
SCOATNet                                       & { 62.06±7.32}           & { 45.40±7.85}           & { 47.93±8.70}           & { 99.77±0.12}           & { 83.94±6.96}           \\
nCoVSegNet                                     & { 75.35 ±3.21}          & { 60.55±4.12}           & { 68.16±5.74}           & { 99.72±0.08}           & { 84.73±2.97}           \\
PDEAtt-Unet                                    & { 68.74±8.55}           & { 46.27±9.00}           & { 64.84±12.10}          & { 99.09±0.24}           & { 61.57±6.87}           \\
{ D-TrAttUnet}             & { 74.44±2.38}           & { 59.34±3.13}           & { 67.97±4.96}           & { 99.68±0.07}           & { 61.74±2.51}           \\ \hline
{ Unet(Hybrid Loss)}       & { 41.82±13.56}          & { 27.44±11.71}          & { 30.75±12.01}           & { 99.69±0.2}            & { 69.89±16.77}          \\
{ UNet++(Hybrid Loss)}     & { 51.37±1.72}           & { 34.58±1.57}           & { 57.72±14.52}          & { 98.52±1.01}           & { 51.74±14.16}          \\
{ Att-Unet(Hybrid Loss)}   & { 50.05±9.74}           & { 33.94±8.65}           & { 38.04±9.12}           & { 99.72±0.11}           & { 74.64±10.76}          \\
{ SCOATNet(Hybrid Loss)}   & { 54.71±18.95}          & { 24.23±8.66}           & { 48.57±23.58}          & { 99.57±0.30}           & { 70.63±10.88}          \\
{ nCoVSegNet(Hybrid Loss)} & { 75.57±2.73}          & { 60.81±3.56}          & { 72.83±4.05}          & { 99.55±0.12}          & { 78.77±4.44}          \\
{ \textbf{CAD-Unet}}       & { \textbf{79.82±1.89}}  & { \textbf{64.46±2.64}}  & { \textbf{78.73±3.81}}  & { \textbf{99.58±0.10}}  & { \textbf{81.18±3.70}}  \\ \hline
{ nnUnet}                  & { 77.95±2.60}          & { 63.92±3.47}          & { 68.17±3.25}          & { 99.85±0.02}          & { 91.06±1.45}          \\
{ \textbf{nnCAD-Unet}}     & { \textbf{77.15±1.07}} & { \textbf{62.81±1.41}} & { \textbf{64.43±1.58}} & { \textbf{99.88±0.01}} & { \textbf{96.17±0.35}} \\ \hline
\end{tabular}
\caption{Comparative experimental results of binary segmentation task on dataset 2.}
\label{tab:2}
\end{table*}

\begin{table*}[h]
\centering
\begin{tabular}{cccccc}
\hline
Model                                         & F1 Score                                    & IoU                                         & Recall                                      & Spec                                        & Prec                                        \\ \hline
Unet                                          & { 74.38±2.77}           & { 57.22±4.46}           & { 66.59±7.70}           & { 99.80±0.03}           & { 79.74±5.14}           \\
UNet++                                        & { 75.99±1.18}           & { 62.08±3.39}           & { 73.95±3.29}           & { 99.77±0.02}           & { 79.37±2.40}           \\
Att-Unet                                      & { 76.64±0.92}           & { 62.09±0.99}           & { 72.86±1.55}           & { 99.84±0.07}           & { 83.09±3.16}           \\
InfNet                                        & { 74.67±0.66}           & { 59.58±0.84}           & { 80.02±1.54}           & { 99.59±0.04}           & { 70.04±1.90}           \\
SCOATNet                                      & { 75.33±1.93}           & { 60.47±2.46}           & { 73.35±4.16}           & { 99.75±0.06}           & { 77.72±3.60}           \\
nCoVSegNet                                    & { 72.92±1.34}           & { 57.40±1.66}           & { 74.94±3.95}           & { 99.62±0.06}           & { 70.34±2.62}           \\
PDEAtt-Unet                                   & { 74.32±0.87}           & { 59.15±1.12}           & { 84.82±1.22}           & { 99.49±0.03}           & { 66.16±1.34}           \\
{ D-TrAttUnet}            & { 71.15±2.00}           & { 55.26±2.41}           & { 81.74±2.29}            & { 99.44±0.08}           & { 63.08±3.24}           \\ \hline
{ Unet(Hybrid Loss)}       & { 74.35±2.71}           & { 59.24±3.38}           & { 68.19±4.36}           & { 99.82±0.02}           & { 81.93±1.70}           \\
{ UNet++(Hybrid Loss)}     & { 75.36±0.70}           & { 60.47±0.90}           & { 83.58±1.33}           & { 99.56±0.02}           & { 68.62±1.04}           \\
{ Att-Unet(Hybrid Loss)}   & { 76.36±1.30}           & { 62.04±1.77}           & { 72.58±2.80}           & { 99.80±0.02}           & { 81.10±1.57}           \\
{ SCOATNet(Hybrid Loss)}   & { 71.72±1.28}           & { 55.93±1.56}           & { 75.38±4.55}           & { 99.59±0.09}           & { 68.76±3.91}           \\
{ nCoVSegNet(Hybrid Loss)} & { 72.97±1.07}           & { 57.46±1.34}           & { 75.89±3.59}           & { 99.62±0.06}           & { 70.48±2.77}           \\
{ \textbf{CAD-Unet}}      & { \textbf{77.97±0.72}}  & { \textbf{63.90±0.98}}  & { \textbf{82.98±1.38}}  & { \textbf{99.65±0.04}}  & { \textbf{73.58±2.25}}  \\ \hline
{ nnUnet}                 & { 82.86±0.21}          & { 70.81±0.29}          & { 84.67±0.65}          & { 99.85±0.01}          & { 81.48±0.61}          \\
{ \textbf{nnCAD-Unet}}    & { \textbf{84.01±0.18}} & { \textbf{72.45±0.27}} & { \textbf{82.16±0.47}} & { \textbf{99.90±0.00}} & { \textbf{86.03±0.46}} \\ \hline
\end{tabular}
\caption{Comparative experimental results of binary segmentation task on dataset 3.}
\label{tab:3}
\end{table*}

\begin{table*}[]
\centering
\begin{tabular}{cccccc}
\hline
Model                                         & F1 Score                                    & IoU                                         & Sens                                        & Spec                                        & Prec                                        \\ \hline
Unet                                          & { 71.61±1.43}           & { 55.79±1.73}           & { 63.36±1.52}           & { 99.83±0.03}           & { 82.41±3.31}           \\
UNet++                                        & { 78.94±0.71}           & { 65.22±0.96}           & { 75.94±1.17}           & { 99.80±0.02}           & { 82.22±1.75}           \\
Att-Unet                                      & { 77.20±0.57}           & { 64.95±0.80}           & { 77.18±1.01}           & { 99.81±0.14}           & { 83.49±0.99}           \\
InfNet                                        & { 70.92±0.53}           & { 54.95±0.64}           & { 71.53±1.33}           & { 99.63±0.02}           & { 70.36±1.37}           \\
SCOATNet                                      & { 79.68±0.32}           & { 66.23±0.45}           & { 76.77±1.41}           & { 99.80±0.02}           & { 82.85±1.13}           \\
nCoVSegNet                                    & { 76.60±0.32}           & { 62.08±0.43}           & { 80.07±0.87}           & { 99.63±0.01}           & { 72.61±0.97}           \\
PDEAtt-Unet                                   & { 77.76±0.48}           & { 62.29±0.63}           & { 83.89±3.03}           & { 99.58±0.06}           & { 70.91±2.88}           \\
{ D-TrAttUnet}            & { 77.16±0.41}           & { 62.82±0.54}            & { 86.12±2.73}           & { 99.55±0.04}           & { 69.97±1.49}           \\ \hline
{ Unet(Hybrid Loss)}       & { 76.76±1.41}           & { 62.31±1.86}           & { 70.96±2.88}           & { 99.83±0.01}           & { 83.71±1.05}           \\
{ UNet++(Hybrid Loss)}     & { 77.38±0.57}           & { 63.11±0.75}           & { 84.68±1.28}           & { 99.59±0.03}           & { 71.27±1.57}           \\
{ Att-Unet(Hybrid Loss)}   & { 77.55±0.56}           & { 64.44±0.78}           & { 77.97±0.97}           & { 99.78±0.09}           & { 82.37±1.98}           \\
{ SCOATNet(Hybrid Loss)}   & { 78.92±0.46}           & { 65.18±0.64}           & { 82.85±1.30}           & { 99.67±0.03}           & { 75.39±1.85}           \\
{ nCoVSegNet(Hybrid Loss)} & { 76.78±0.29}           & { 62.32±0.39}           & { 80.79±0.93}           & { 99.64±0.02}           & { 73.18±1.11}           \\
{ \textbf{CAD-Unet}}      & { \textbf{80.83±0.20}}  & { \textbf{67.83±0.28}}  & { \textbf{86.17±1.72}}  & { \textbf{99.69±0.02}}  & { \textbf{76.71±1.06}}  \\ \hline
{ nnUnet}                 & { 85.16±0.17}          & { 74.16±0.26}          & { 81.85±0.18}          & { 99.88±0.01}          & { 88.75±0.49}          \\
{ \textbf{nnCAD-Unet}}    & { \textbf{85.56±0.20}} & { \textbf{74.77±0.30}} & { \textbf{82.58±0.56}} & { \textbf{99.88±0.01}} & { \textbf{88.78±0.73}} \\ \hline
\end{tabular}
\caption{Comparative experimental results of binary segmentation task on dataset 4.}
\label{tab:4}
\end{table*}
\subsubsection{Multi-class segmentation task}
For the multi-class segmentation task, we merge dataset 1 and dataset 2 into a single multi-class segmentation dataset. On this combined dataset, we compare the proposed model with U-Net, Att-UNet, U-Net++, CopleNet, AnamNet, SCOAT-Net, D-TrAttUnet, and nnUNet. The experimental results, as shown in Table 6, demonstrate that the proposed model also performs well on this task. Specifically, compared to D-TrAttUnet, the best-performing non-nnUNet-based architecture, the proposed model achieves a slight improvement of 0.1 in the F1 score for the GGO class and a more significant improvement of 1.67 for the more challenging CON class. After integrating the proposed model into nnUNet, although the F1 score for the GGO class decreases by 0.26, it achieves a substantial improvement of 2.26 for the CON class, along with a lower standard deviation, indicating greater model stability.

The experimental results collectively demonstrate that the proposed model achieves competitive performance in both binary and multi-class segmentation tasks. This is attributed to the capsule network's ability to identify complex lesion postures and the ingenious pathway coupling mechanism. The capsule network effectively captures the unique geometric and spatial features of lesions, while the pathway coupling mechanism facilitates efficient information fusion. Their synergy leads to accurate lesion detection, as evidenced by the experimental results across four datasets in both tasks. These elements are crucial for enhancing the segmentation of COVID-19 lung CT lesions, offering a reliable approach in medical image analysis.

\begin{table*}[t]
\centering
\begin{tabular}{ccccc}
\hline
\multirow{2}{*}{Model} & \multicolumn{2}{c}{GGO}                     & \multicolumn{2}{c}{CON}                     \\ \cline{2-5} 
                       & F1 score                 & IoU                  & F1 score                 & IoU                  \\ \hline
Unet                   & 49.75±8.52           & 33.54±7.58           & 43.23±4.98           & 27.70±3.91           \\
Att-Unet               & 69.06±0.88           & 53.46±1.04           & 54.61±1.62           & 37.58±1.53           \\
Unet++                 & 65.69±1.29           & 48.92±14.2           & 31.31±6.67           & 18.75±4.73           \\
CopleNet               & 60.44±1.54           & 43.33±1.61           & 29.70±10.29           & 17.90±7.52            \\
AnamNet                & 65.10±3.56            & 48.36±3.82           & 31.97±6.12           & 19.18±4.36           \\
SCOATNET               & 65.77±3.28           & 49.09±3.56           & 43.52±1.67           & 27.83±1.38           \\
D-TrAttUnet            & 70.61±1.01           & 54.58±1.21           & 57.94±2.30           & 40.82±3.66           \\
\textbf{CAD-Unet}      & \textbf{70.71±0.44}  & \textbf{54.69±0.52}  & \textbf{59.61±0.74}  & \textbf{42.47±0.75}  \\ \hline
nnUnet                 & 73.39±1.02           & 57.97±1.26           & 60.30±2.20           & 43.19±2.23           \\
\textbf{nnCAD-Unet}    & \textbf{73.13±0.22} & \textbf{57.65±0.27} & \textbf{62.56±0.85} & \textbf{45.53±0.90} \\ \hline
\end{tabular}
\caption{Comparative experimental results of multi-class segmentation task on dataset 1 and dataset 2.}
\label{tab:5}
\end{table*}

\begin{table}[]
\begin{tabular}{lccc}
\hline
{Model}                  & \multicolumn{1}{c}{{Dataset 2}} & \multicolumn{1}{c}{{Dataset 3}} & \multicolumn{1}{c}{{Dataset 4}} \\ \hline
{Unet}                   & {0.009}                         & {0.009}                         & {0.009}                         \\
{UNet++}                 & {0.009}                         & {0.047}                         & {0.009}                         \\
{Att-Unet}               & {0.009}                         & {0.047}                         & {0.009}                         \\
{InfNet}                 & {0.009}                         & {0.009}                         & {0.009}                         \\
{SCOATNet}               & {0.009}                         & {0.028}                         & {0.009}                         \\
{nCoVSegNet}             & {0.047}                         & {0.009}                         & {0.009}                         \\
{PDEAtt-Unet}            & {0.009}                         & {0.009}                         & {0.009}                         \\
{D-TrAttUnet}            & {0.009}                         & {0.009}                         & {0.009}                         \\
{Unet(Hybrid Loss)}       & {0.009}                         & {0.009}                         & {0.009}                         \\
{UNet++(Hybrid Loss)}     & {0.009}                         & {0.009}                         & {0.009}                         \\
{Att-Unet(Hybrid Loss)}   & {0.009}                         & {0.028}                         & {0.009}                         \\
{SCOATNet(Hybrid Loss)}   & {0.009}                         & {0.009}                         & {0.009}                         \\
{nCoVSegNet(Hybrid Loss)} & {0.047}                         & {0.009}                         & {0.009}                         \\ \hline
\end{tabular}
\caption{P-values for binary segmentation across datasets.}
\label{tab:d}
\end{table}

\begin{table}[]
\centering
\begin{tabular}{lcc}
\hline
{Model}       & \multicolumn{1}{l}{{GGO}} & \multicolumn{1}{l}{{CON}} \\ \hline
{Unet}        & {0.009}                   & {0.009}                   \\
{Att-Unet}    & {0.047}                   & {0.009}                   \\
{Unet++}      & {0.009}                   & {0.009}                   \\
{CopleNet}    & {0.009}                   & {0.009}                   \\
{AnamNet}     & {0.009}                   & {0.009}                   \\
{SCOATNET}    & {0.009}                   & {0.009}                   \\
{D-TrAttUnet} & {0.32}                    & {0.045}                   \\ \hline
\end{tabular}
\caption{P-values for multi-class segmentation.}
\label{tab:e}
\end{table}

\begin{table*}[h]
\centering
\begin{tabular}{c|ccc|ccc|ccc}
\hline
\multirow{2}{*}{Architecture} & \multicolumn{3}{c|}{Ablation} & \multicolumn{3}{c|}{Dataset 2}                         & \multicolumn{3}{c}{Dataset 3}                                   \\ \cline{2-10} 
                              & CAP       & AG      & DD      & F1 score       & IoU                 & Recall                 & F1 score                & IoU                 & Recall                 \\ \hline
ResUnet(baseline)             & $\times$         & $\times$       & $\times$       & 53.14±8.45 & 36.62±7.51          & 43.84±11.08         & 74.99±1.76          & 75.99±1.76          & 78.40±3.94           \\
AttUnet(baseline)             & $\times$         & $\checkmark$       & $\times$       & 63.63±8.74 & 46.77±6.02          & 48.96±7.37          & 76.25±1.05          & 61.09±0.99          & 71.86±1.55          \\
Cap                           & $\checkmark$         & $\times$       & $\times$       & 61.74±7.00 & 45.02±7.27          & 51.74±10.63         & 75.39±0.61          & 60.31±0.79          & 77.53±3.03          \\
AG\&DD                        & $\times$         & $\checkmark$       & $\checkmark$       & 71.06±5.90 & 55.41±6.69           & 63.68±5.35          & 77.34±0.63          & 63.05±0.85          & 79.60±1.48            \\
Cap\&DD                       & $\checkmark$         & $\times$       & $\checkmark$       & 73.12±3.80 & 57.77±4.72          & 64.32±4.74          & 77.24±1.29          & 62.94±1.72          & 81.23±3.32          \\
Cap\&AG                       & $\checkmark$         & $\checkmark$       & $\times$       & 64.97±7.13 & 48.50±7.32          & 54.45±6.65          & 74.93±0.67          & 59.91±0.86          & 80.80±2.62          \\
\textbf{Cap\&AG\&DD}                   & $\checkmark$         & $\checkmark$       & $\checkmark$       & \textbf{79.82±1.89} & \textbf{64.46±2.64} & \textbf{78.73±3.81} & \textbf{77.97±0.72} & \textbf{63.90±0.98} & \textbf{82.98±1.38} \\ \hline
\end{tabular}
\begin{tabular}{c|ccc|ccc}
\hline
\multirow{2}{*}{Architecture} & \multicolumn{3}{c|}{Ablation} & \multicolumn{3}{c}{Dataset 4}                                   \\ \cline{2-7} 
                              & CAP       & AG      & DD      & F1 score                & IoU                 & Recall                 \\ \hline
ResUnet(baseline)             & $\times$         & $\times$       & $\times$       & 76.79±0.41          & 65.00±0.56          & 83.23±0.87          \\
AttUnet(baseline)             & $\times$         & $\checkmark$       & $\times$       & 77.20±0.57          & 66.95±0.80          & 77.18±1.01          \\
Cap                           & $\checkmark$         & $\times$       & $\times$       & 80.49±0.32          & 67.36±0.45          & 86.17±0.53          \\
AG\&DD                        & $\times$         & $\checkmark$       & $\checkmark$       & 80.43±0.35          & 66.00±0.48          & 85.62±1.97          \\
Cap\&DD                       & $\checkmark$         & $\times$       & $\checkmark$       & 80.51±0.40          & 65.89±0.55          & 85.25±1.77          \\
Cap\&AG                       & $\checkmark$         & $\checkmark$       & $\times$       & 80.21±0.17          & 65.58±0.23          & 84.28±1.08          \\
\textbf{Cap\&AG\&DD}                   & $\checkmark$         & $\checkmark$       & $\checkmark$       & \textbf{80.83±0.20} & \textbf{67.83±0.28} & \textbf{86.17±1.72} \\ \hline
\end{tabular}
\caption{The results of ablation experiments for binary segmentation tasks. The impact of three modules, namely Cap, AG, and DD, on the overall architectural performance in binary segmentation is validated on dataset 2, dataset 3, and dataset 4.}
\label{tab:6}
\end{table*}

% \usepackage{multirow}
\begin{table*}[h]
\centering
\begin{tabular}{c|ccc|ccc|ccc}
\hline
\multirow{2}{*}{Architecture} & \multicolumn{3}{c|}{Ablation}              & \multicolumn{3}{c|}{GGO}                                        & \multicolumn{3}{c}{CON}                                         \\ \cline{2-10} 
                              & CAP          & AG           & DD           & F1 score                & IoU                 & Recall                 & F1 score                & IoU                 & Recall                 \\ \hline
ResUnet(baseline)             & $\times$     & $\times$     & $\times$     & 68.41±1.89        & 52.02±1.41        & 61.27±2.42          & 54.39±2.05        & 37.89±1.95        & 59.61±5.51          \\
AttUnet(baseline)             & $\times$     & $\checkmark$ & $\times$     & 69.66±0.88          & 53.46±1.04          & 62.05±1.33          & 54.61±1.62          & 37.58±1.53          & 60.18±2.62          \\
Cap                           & $\checkmark$ & $\times$     & $\times$     & 67.90±2.55          & 51.45±2.88          & 59.40±4.73          & 55.28±1.95          & 38.22±1.86          & 61.31±1.90          \\
AG\&DD                        & $\times$     & $\checkmark$ & $\checkmark$ & 67.51±3.94          & 51.09±4.44          & 58.99±5.87          & 56.52±1.82          & 39.41±1.79          & 59.47±4.59          \\
Cap\&DD                       & $\checkmark$ & $\times$     & $\checkmark$ & 68.09±2.53          & 51.67±2.86          & 58.97±4.58          & 56.37±0.58          & 39.25±0.56          & 62.86±2.65          \\
Cap\&AG                       & $\checkmark$ & $\checkmark$ & $\times$     & 69.84±1.43          & 53.67±1.66          & 63.53±1.34          & 55.89±1.50          & 38.80±1.44          & 58.47±0.80          \\
Cap\&AG\&DD                   & $\checkmark$ & $\checkmark$ & $\checkmark$ & \textbf{70.71±0.44} & \textbf{54.69±0.52} & \textbf{63.34±0.98} & \textbf{59.61±0.74} & \textbf{42.47±0.75} & \textbf{64.59±2.14} \\ \hline
\end{tabular}
\caption{The results of ablation experiments for multi-class segmentation tasks. The influence of three modules, namely Cap, AG, and DD, on the overall architectural performance in multi-class segmentation is verified on the combined datasets of dataset 1 and dataset 2.}
\label{tab:7}
\end{table*}

\begin{figure*}[h] % 开始图片
\centering % 图片居中
\includegraphics[scale=0.7]{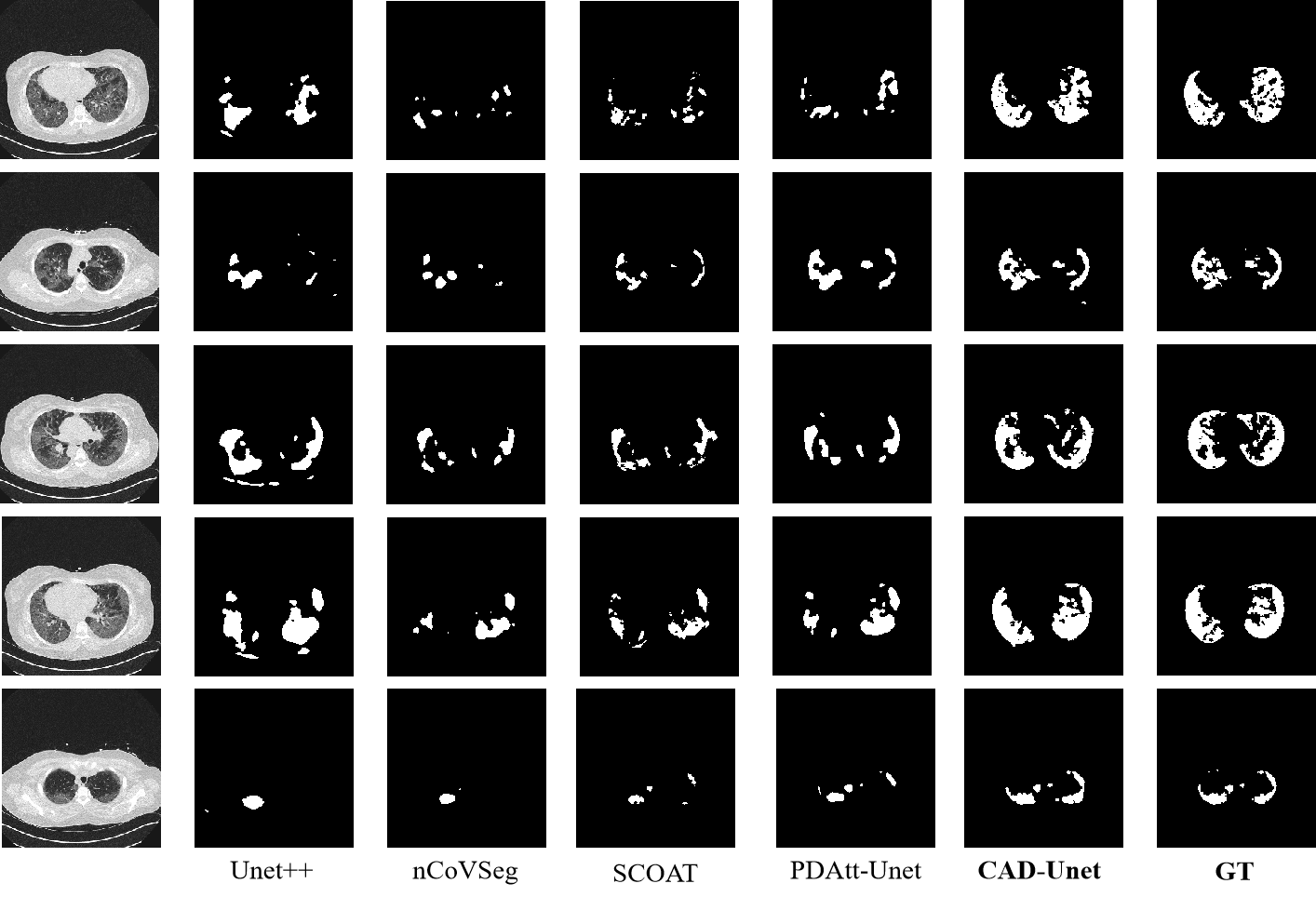} %[设置图像大小]{图片路径(需要在本文件下的路径)}
\caption{Visual comparison of a binary segmentation model trained with different segmentation architectures for binary COVID-19 segmentation using dataset 2, dataset 3 and dataset 4.} % 图像描述
\label{fig:7} % 给图片标记
\end{figure*}% 结束图片

\begin{figure*}[h] % 开始图片
\centering % 图片居中
\includegraphics[scale=0.5]{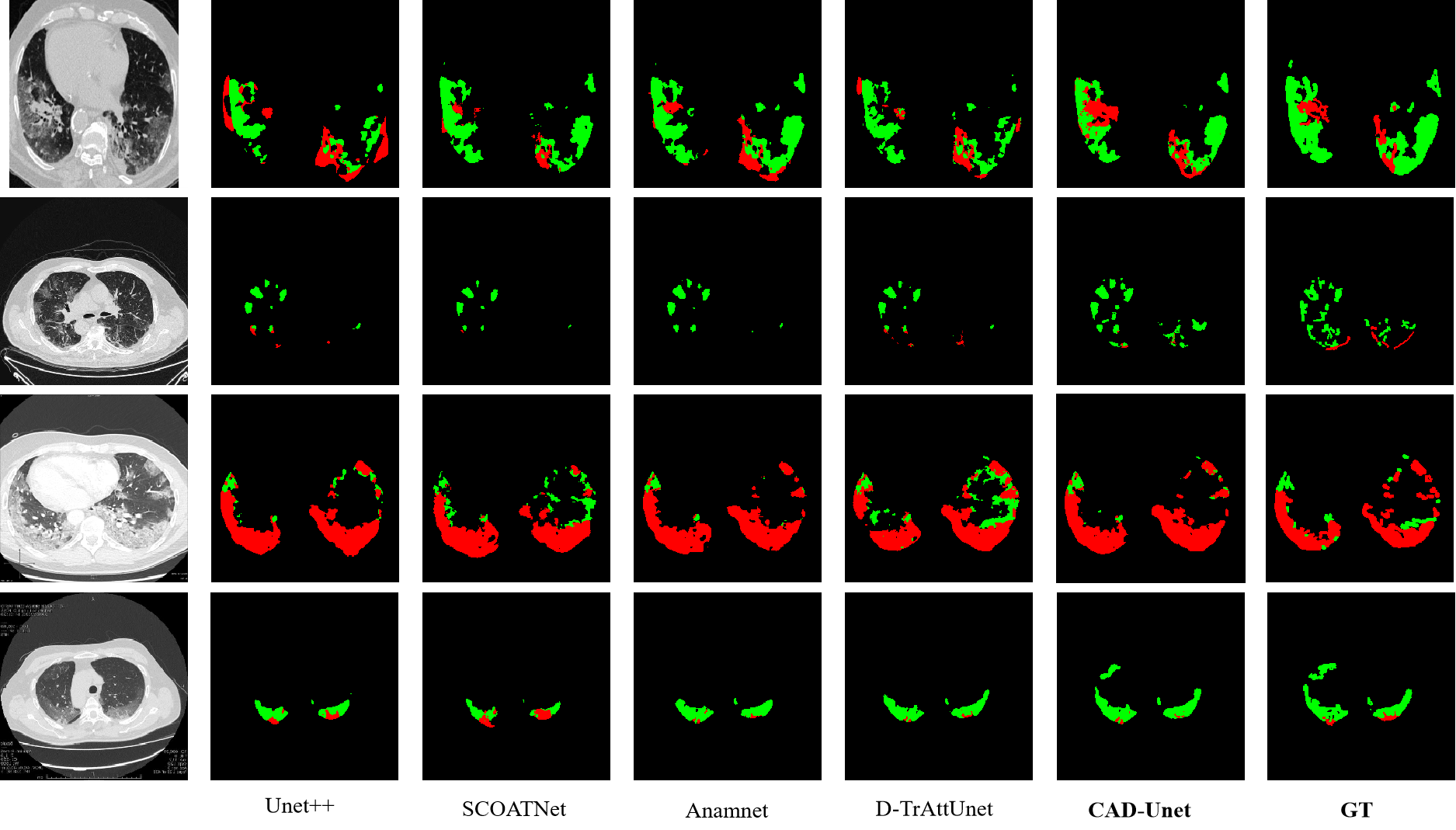} %[设置图像大小]{图片路径(需要在本文件下的路径)}
\caption{ Visual comparison of a segmentation model trained with different segmentation architectures for multi-class (No-infection, GGO and CON) COVID-19 infection segmentation using dataset 1 and dataset 2. GGO is represented in green and CON in red.} % 图像描述
\label{fig:8} % 给图片标记
\end{figure*}% 结束图片

\subsection{Statistical tests}
To test the statistical significance of the proposed model's performance, we conduct statistical tests on the F1 scores for both tasks. Given the non-normal distribution of performance metrics typical in deep learning, we adopt the Wilcoxon rank-sum test. The F1 scores of the proposed model and compared models are treated as independent samples, and p-values are computed for each pairwise comparison. A p-value $<$ 0.05 indicates that the observed differences are unlikely to be due to chance, providing evidence against the null hypothesis and supporting the superiority of the proposed model. The p-values for both tasks are summarized in Tables \ref{tab:d} and \ref{tab:e}.

For the binary segmentation task, across all three datasets, the p-values between the proposed model and all compared models are below 0.05, indicating that the performance improvements of the proposed model are statistically significant. 

For the multi-class segmentation task, the proposed model demonstrates significant improvements over most baseline models.Notably, in the GGO classification, the p-value of 0.32 when compared to the D-TrAttUnet model indicates no statistically significant difference between the two models. However, in the CON classification, the p-value of 0.045 suggests a significant performance difference. Additionally, the proposed model exhibits a lower standard deviation in the GGO classification, highlighting its stability. Given the sensitivity of medical image segmentation tasks to stability, these results suggest that the proposed model remains a strong candidate despite the lack of statistical significance in certain cases.

\subsection{Ablation study}

To evaluate the effectiveness of CAD-Unet and its individual components, ablation experiments are conducted across four datasets and two tasks. Referring to the viewpoints in \cite{10.1007/978-3-031-72114-4_47}, and to avoid baseline-related pitfalls, we employ ResUnet, a U-Net with residual connections, and AttUnet as baseline models. The results for binary segmentation on datasets 2, 3, and 4 are summarized in Table \ref{tab:6}, while those for multi-class segmentation on datasets 1 and 2 are presented in Table \ref{tab:7}. These experiments assess the contributions of the capsule layer, dual decoders, and attention gate.

In the binary segmentation task, we first evaluate the impact of the CAD-Unet module on the baseline ResUnet model. As shown in Table \ref{tab:6}, incorporating the attention gate and capsule module individually leads to improvements in F1 scores by 10.49, 1.26, 0.41 and 8.6, 0.4, 3.7 for datasets 2, 3,  4, respectively. The enhancements in other metrics also vary, indicating that both the capsule module and attention gate contribute to performance improvement. Next, we examine the combined effects of the capsule, dual decoders, and attention gate on binary segmentation performance. Disabling the capsule results in F1 score reductions of 8.76, 0.63, and 0.4 on the three datasets, highlighting the importance of the capsule layer in capturing hierarchical spatial relationships and enhancing feature representation. The deactivation of the attention gate causes F1 score reductions of 6.7, 0.73, and 0.32, while disabling the dual decoders leads to even larger reductions of 14.85, 3.04, and 0.62. These results suggest that the attention gate and dual decoders also significantly contribute to performance improvements in binary segmentation tasks. The dual decoders help the network focus on the lung region, while the attention gate design enables the model to concentrate on critical spatial information, enhancing the recognition of subtle features.

In the multi-class segmentation task, similar ablation experiments are conducted. As shown in Table \ref{tab:7}, incorporating the attention gate module into the baseline model results in F1 score improvements of 1.25 and 0.22 percentage points for GGO and CON classifications, respectively. Incorporating the capsule module into the ResUnet model results in an F1 score decrease of 0.51 for GGO classification but an improvement of 0.89 for CON classification, which indicates that the standalone capsule module performs poorly for GGO classification but works well for CON classification. However, adding both the attention gate and capsule module to the baseline model results in improvements of 1.43 and 1.5 for GGO and CON classifications, respectively, suggesting that the interaction between the two modules complements each other’s weaknesses, leading to better multi-class feature recognition. When compared with the complete architecture, disabling the capsule module leads to F1 score reductions of 3.2 and 3.09 for GGO and CON classifications, respectively, emphasizing the capsule's role in capturing hierarchical relationships between categories. The effects of the attention gate and dual decoders on multi-class segmentation are also evaluated. Disabling the attention gate results in F1 score reductions of 2.62 and 3.24 for GGO and CON classifications, while disabling the dual decoders leads to F1 score reductions of 0.87 and 3.72 for the same classifications. These findings indicate that both the attention gate and dual decoders play significant roles in enhancing the performance of the model in multi-class segmentation tasks.

In summary, the ablation experiments demonstrate that the capsule module, dual decoders, and attention gate are indispensable components of CAD-Unet, collectively contributing to the model's outstanding performance in COVID-19 lung infections segmentation tasks. These results validate the effectiveness of our proposed architecture and provide valuable insights for future research in deep learning for medical image segmentation.
% \usepackage{multirow}
% \usepackage{multirow}
% \usepackage{multirow}

\section{Discussion}
\subsection{Qualitative analysis}
To visually demonstrate the efficacy of our proposed model in addressing both binary and multi-class segmentation tasks for COVID-19 lung CT lesion identification, we conduct a qualitative examination of the segmentation outcomes for both scenarios. Fig.\ref{fig:7} provides a visual representation of partially predicted masks alongside the ground truth (GT) for the binary segmentation task, derived from three distinct datasets. The proposed model is compared with Unet++, nCoVSeg, SCOAT-Net, and PDAtt-Unet models. The results clearly demonstrate the superior performance of the predicted masks by our model, surpassing the other four models in terms of accurately encompassing the entire lesion region and exhibiting more precise and well-defined segmentation boundaries.\par
This demonstrates the capability of the model in effectively addressing binary segmentation challenges associated with COVID-19 lung lesions. The predicted masks produce by our model consistently demonstrate a propensity to encapsulate intricate lesion shapes while displaying sharper and more precise boundaries compared to the alternative models.\par
Fig.\ref{fig:8} illustrates the visualization of partially predicted masks and GT for the multi-class segmentation task from two datasets, using our proposed model along with Unet++, SCOAT-Net, Anamnet, and D-TrAttUnet models. Comparing the predicted masks and GT of other models, the first row of images shows that the proposed model captures more complete CON class masks. The second row demonstrates the superior segmentation performance of our model on small target regions. The third row indicates that our model obtains more accurate target shape information, while the fourth row shows that we achieve more complete target segmentation masks in the GGO class.\par
In conclusion, our comprehensive qualitative analysis provides evidence suggesting that the proposed model performs favorably compared to current alternatives in addressing binary and multi-class segmentation challenges inherent to COVID-19 lung CT lesion identification. The visual evidence presented here indicates that the model effectively captures intricate lesion morphology and delineates their boundaries with accuracy. Collectively, these advancements may enhance medical image analysis, potentially supporting medical practitioners in their efforts to manage the evolving landscape of the COVID-19 pandemic.

\begin{table*}[]
\centering
\begin{tabular}{llll}
\hline
Model             & Computational cost & Number of parameters & Inference time \\ \hline
Unet              & 474.48 GFLOPs      & 7.96 M               & 2.7 ms         \\
UNet++            & 1698.59 GFLOPs     & 36.62 M              & 3.3 ms         \\
Att-Unet          & 211.46 GFLOPs      & 10.29 M              & 4.3 ms         \\
InfNet            & 90.69 GFLOPs       & 31.07 M              & 16.5 ms        \\
AnamNet           & 311.57 GFLOPs      & 4.63 M               & 3.9 ms         \\
SCOATNet          & 481.11 GFLOPs      & 10.21 M              & 7.7 ms         \\
nCoVSegNet        & 636.72 GFLOPs      & 168.17 M             & 13.3 ms        \\
PDEAtt-Unet       & 331.26 GFLOPs      & 14.44 M              & 8.9 ms         \\
D-TrAttUnet       & 661.54 GFLOPs      & 104.16 M             & 13.0 ms        \\
\textbf{CAD-Unet} & 362.91 GFLOPs      & 15.04 M              & 23.4 ms        \\ \hline
\end{tabular}
\caption{Comparison of computational cost, model parameters, and inference time (Batch Size = 16)}
\label{tab:c}
\end{table*}
\subsection{Computational efficiency analysis}
Efficient model design is critical for real-world clinical applications of medical image segmentation. Table \ref{tab:c} presents the computational cost, parameter count, and inference time for our proposed model and several baseline methods, where computations are performed with a batch size of 16 images.

Compared to the Transformer-based D-TrAttUnet (661.54 GFLOPs, 104.16M parameters), our Capsule Network-based model achieves competitive segmentation performance with significantly lower computational cost (362.91 GFLOPs) and model size (15.04M parameters). This demonstrates that capsule networks can achieve efficient feature representation with relatively low computational cost and parameter count, making them a promising alternative to Transformer-based architectures for medical image segmentation.

In terms of inference speed, our model achieves a batch inference time of 23.4 ms for 16 images, corresponding to an average of 1.4 ms per image, which is suitable for real-time medical diagnosis. However, compared to other models such as UNet (2.698 ms), UNet++ (3.300 ms), and Att-UNet (4.312 ms), our inference time is relatively longer. The increased latency may be attributed to the iterative routing mechanisms in capsule networks, which enhance feature representation but introduce additional computational overhead. Nevertheless, given its strong segmentation performance and efficient feature encoding, our model remains a viable choice for medical image analysis tasks that require both accuracy and robustness.

\section{Conclusions}
The outbreak of COVID-19 has presented significant challenges to global health.    Lung CT imaging plays a crucial role in the early diagnosis and treatment of COVID-19 patients. This paper aims to develop an efficient and accurate lung CT lesion segmentation method to assist clinicians in quickly obtaining lesion information for more precise diagnosis and treatment. Our proposed CAD-Unet architecture combines the Unet encoder pathway and capsule encoder pathway in parallel to extract richer local, global, and long-range dependent features. Furthermore, to address the limitations of the BCE loss function in segmenting infection edges, we incorporate an edge loss component. \par To assess the effectiveness of the proposed method, we conduct evaluations on two tasks: binary segmentation and multi-class segmentation. In addition, we compare our method with three baselines and five state-of-the-art architectures for COVID-19 segmentation. The experimental results demonstrate that our method outperforms all the mentioned architectures in common evaluation metrics for both binary and multi-class segmentation tasks. Furthermore, we perform  ablation experiments to validate the effectiveness of the components in our proposed model. In future work, we will explore the potential of combining capsule networks with traditional segmentation models and applying them to a wider range of medical image segmentation tasks.

\section*{Acknowledgments}
The authors would like to thank the Editor and the anonymous reviewers for their constructive comments to improve this manuscript. This work was supported by the National Natural Science Foundation of China (Grant Nos. 62463025, 62462051), by the Natural Science Foundation of Ningxia (Grant No. 2024AAC03038), and by the Youth Top Talent Training Program of Ningxia 2022 (Grant No. 2022012).

\section*{Appendix.}
\begin{figure*}[h] % 开始图片
\centering % 图片居中
\includegraphics[scale=0.39]{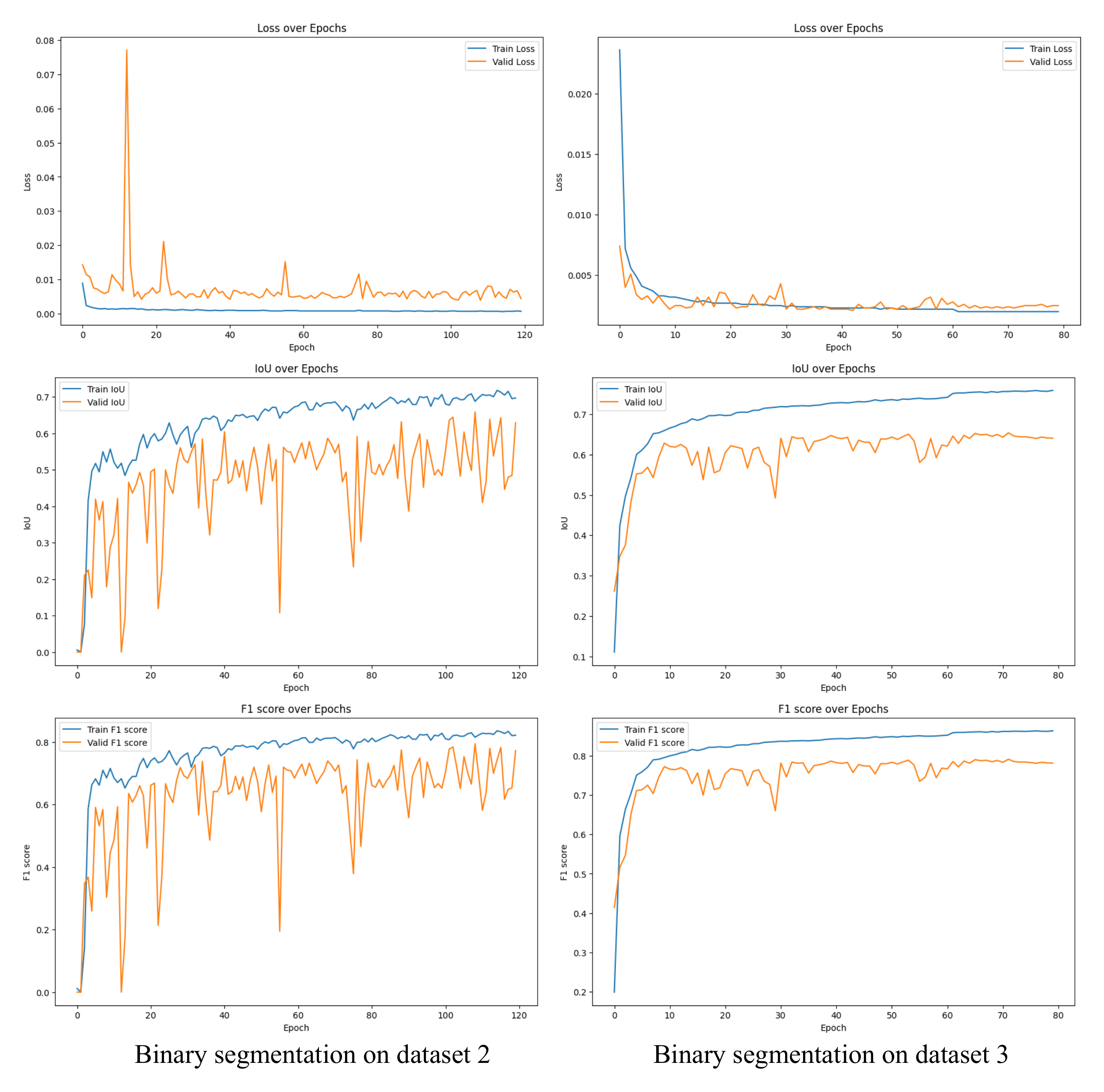}
\caption{Learning curves (1)} % 图像描述
\label{fig:a1} % 给图片标记
\end{figure*}

\begin{figure*}[h] % 开始图片
\centering % 图片居中
\includegraphics[scale=0.39]{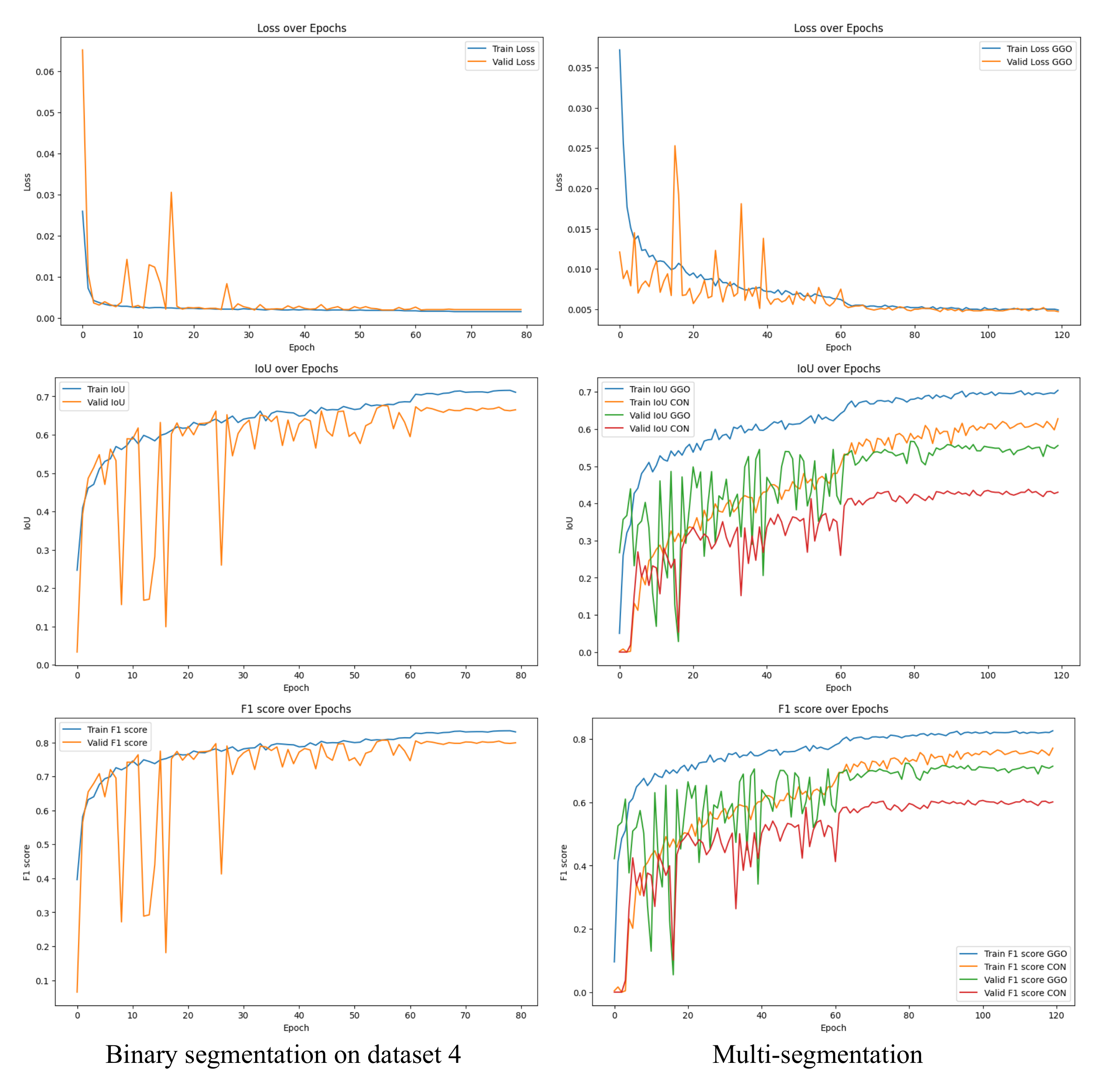}
\caption{Learning curves (2)} % 图像描述
\label{fig:a2} % 给图片标记
\end{figure*}

In the appendix, we attach the learning curves for binary and multi-class segmentation tasks in Fig.\ref{fig:a1} and Fig.\ref{fig:a2}. The curves show the changes in loss, IoU, and F1 score during training, which help evaluate the model's performance. 

% 结束图片

% 结束图片

%\section{\itshape Reference style}

%Text: All citations in the text should refer to:
%\begin{enumerate}
%\item Single author: the author's name (without initials, unless there
%is ambiguity) and the year of publication;
%\item Two authors: both authors' names and the year of publication;
%\item Three or more authors: first author's name followed by `et al.'
%and the year of publication.
%\end{enumerate}
%Citations may be made directly (or parenthetically). Groups of
%references should be listed first alphabetically, then chronologically.

%%Harvard
\bibliographystyle{model2-names.bst}\biboptions{authoryear}
\bibliography{refs}

%\section*{Supplementary Material}

%Supplementary material that may be helpful in the review process should
%be prepared and provided as a separate electronic file. That file can
%then be transformed into PDF format and submitted along with the
%manuscript and graphic files to the appropriate editorial office.

\end{document}